\documentclass[11pt]{article}
\usepackage{palatino}

\usepackage{amsmath}
\usepackage{amssymb}

\renewcommand{\phi}{\varphi}

\newcommand{\secref}[1]{\textsection\ref{chap13:#1}}

\newcommand{\Rule}[2]{          
  \begin{array}{c}
  #1 \\\hline
  #2
  \end{array}}

\newcommand{\mi}[1]{\mathit{#1}}
\newcommand{\ms}[1]{\mathsf{#1}}

\newcommand{\sees}{~\ms{sees}~}
\newcommand{\said}{~\ms{said}~}
\newcommand{\controls}{~\ms{controls}~}
\newcommand{\fresh}{\ms{fresh}}
\newcommand{\believes}{~\ms{believes}~}

\newcommand{\key}[1]{\mathrel{\stackrel{\scriptscriptstyle {#1}}{\leftrightarrow}}}

\newcommand{\secret}[1]{\mathrel{\stackrel{\scriptscriptstyle
{#1}}{\rightleftharpoons}}}

\newenvironment{protocol}[1]
{\begin{equation}\label{#1}\begin{array}{rcccll}}
{\end{array}\end{equation}}

\newcommand{\pmessage}[4]{#1. \quad & #2
  & \hspace{-1ex}\longrightarrow\hspace{-1ex} & #3 & : \quad & #4}

\newenvironment{interaction}[1]
  {\begin{equation}\label{#1}\begin{array}{cccll}}
  {\end{array}\end{equation}}

\newcommand{\imessage}[3]{#1
  & \hspace{-1ex}\longrightarrow\hspace{-1ex} & #2 & : \quad & #3}

\newcommand{\cin}{\,\mathord{\in}\,}

\newcommand{\sencr}[2]{\{#1\}_{#2}} 

\newcommand{\rimp}{\Rightarrow}

\newcommand{\restr}[1]{\rvert_{#1}}

\newcommand{\fby}[2]{\langle #1, #2\rangle}

\newcommand{\cK}{\mathcal{K}}
\newcommand{\sK}{\mathsf{K}}

\newcommand{\qq}[1]{[\hspace{-.3ex}[#1]\hspace{-.3ex}]}

\title{Knowledge and Security}
\author{Riccardo Pucella\thanks{Author's email: \texttt{riccardo@acm.org}. To appear in \textit{Handbook of Logics for Knowledge and Belief}.}}
\date{}

\begin{document}

\maketitle

\begin{abstract}
Epistemic concepts, and in some cases epistemic logic, have been used
in security research to formalize security properties of systems. This
survey illustrates some of these uses by focusing on
confidentiality in the context of cryptographic protocols, and in the
context of multi-level security systems.  
\end{abstract}

\bigskip

\begin{flushright}
\textit{Security: the state of being free from danger or threat.}\\[1ex]
\textit{(New Oxford American Dictionary)}
\end{flushright}

\section{Introduction}\label{chap13:introduction}

A persistent intuition in some quarters of the security
research community says that epistemic logic and, more generally,
epistemic concepts are useful for reasoning about the
security of systems. 
What grounds this intuition is that much work in the field is based on
epistemic concepts---sometimes explicitly, but more often implicitly,
by and large reinventing possible-worlds semantics for knowledge and
belief.\footnote{This chapter assumes from the reader a basic
  knowledge of epistemic logic and its Hintikka-style possible-worlds
  semantics; see \secref{bibnotes} for references.
Furthermore, to simplify the exposition, the term \emph{epistemic} is
used to refer both to knowledge and to belief throughout.
}

Reasoning about the security of systems in practice amounts to
establishing that those systems satisfy various security
  properties. A security property, roughly speaking, is a property of
a system stating that the system is not vulnerable to a particular
threat. 
Threats, in this context, are generally taken to be attacks
by agents intent on subverting the system.

While what might be considered a threat---and therefore what security
properties are meant to protect against---is in the eye of the beholder,
several properties have historically been treated as security
properties:

\begin{itemize}

\item \textbf{Data Confidentiality:} only authorized agents should have access
  to a piece of data; more generally, only authorized agents should be able
  to infer any information about a piece of data.

\item \textbf{Data Integrity:} 
only authorized agents should have access to alter a piece of data.

\item \textbf{Agent Authentication:} an agent should be able to prove her identity to another agent.

\item \textbf{Data Authentication:} an agent should be able to determine the
  source of a piece of data. 

\item \textbf{Anonymity:} the identity of an agent or the source of
  a piece of data should be kept hidden except from authorized agents.

\item \textbf{Message Non-repudiation:} the sender of a message should
  not be able to deny having sent the message.

\end{itemize}
These properties may seem intuitive on a first reading but they
are vague and depend on terms that require
clarification: \emph{secret}, \emph{authorized}, \emph{access
  to a piece of data}, \emph{source}, \emph{identity}.

Epistemic concepts come into play when defining many of the terms that
appear in the statements of security properties.
Indeed, those terms
can often be usefully understood in terms of knowledge:
confidentiality can be read as \emph{no agent except for authorized agents
  can know a piece of information}; authentication as \emph{an agent
  knows the identity of the agent with whom she is interacting}, or
\emph{an agent knows the identity of the agent who sent the
  information};  anonymity as \emph{no one knows the identity of the
  agent who performed a particular action}; and so on. 
While it is not the case that every security property can be
read as an epistemic property, enough of them can to justify studying
them \emph{as} epistemic properties.

Epistemic logic and epistemic concepts play two roles in security
research: 
\begin{itemize}

\item \textbf{Definitional}: they are used to formalize security
  properties and concepts, and provide a clear semantic grounding
  for them. Epistemic logic may be explicitly used as an explanatory
  and definitional language for properties of interest.

\item \textbf{Practical}: they are used to derive verification
  and enforcement techniques for security properties, that is, to
  either establish that a security property is true in a system, or to
  force a security property to be true in a system. 

\end{itemize}

It is fair to say that after nearly three decades of
research, epistemic logic has had several successes on the
definitional front and somewhat fewer on the practical front. 
This is perhaps not surprising. While epistemic logic and epistemic
concepts are well suited for definitions and for describing semantic
models, verification of epistemic properties tends to be expensive,
and tools for the verification of security properties in practice
often approximate epistemic properties using properties that are
easier to check, such as safety properties.\footnote{A safety property
  is a property of the form \emph{a bad state is never reached in any
    execution of the system}, for some definition of \emph{bad
    state}. A safety property can be checked by examining every
  possible execution independently of any other; in contrast,
  checking an epistemic property requires examining every possible
  execution in the context of all other possible executions.}

This chapter illustrates the use of epistemic logic and epistemic
concepts for reasoning about security through the study of a specific
security property, confidentiality.
Not only is confidentiality a prime example of the use of
knowledge to make a security property precise, 
but it has also been studied extensively from several
perspectives.
Moreover, many of the issues arising while studying confidentiality also
arise for other security properties with an epistemic flavor.
 
Confidentiality is explored in two contexts: cryptographic protocols in
\secref{cryptoprotocols}, and multi-level security systems in
\secref{iflow}. Cryptographic protocols are communication protocols
that use cryptography to protect information exchanged between agents in a system. While it may seem simple enough for Alice to
send a confidential message to Betty by encrypting it, 
Alice and Betty need to share a common key for this to work. How is
such a key distributed before communication may take place?  
Most cryptographic protocols involve key creation and distribution,
and these are notoriously difficult to get right.
Key distribution also forces the consideration of authentication as an
additional security property.
Cryptographic protocol analysis is the one field of security research
that has explicitly and extensively used epistemic logic, and the bulk
of this chapter is dedicated to that topic. 

In multi-level security systems, one is generally interested in
confidentiality guarantees even when information is
used or released within the system during a computation. 
The standard example is that of a centralized system where agents have
different security clearances and interact with data with different
security classifications; the desired confidentiality guarantees
ensure that classified data, no matter how it is manipulated by agents
with an appropriate security clearance, never flows to an agent that
does not have an appropriate security clearance. 
Most of the work in this field of security research uses epistemic
concepts implicitly--- the models use possible-worlds definitions of
knowledge, but no epistemic logic is introduced. All reasoning is
semantic reasoning in the models.

Security properties other than confidentiality are briefly discussed
in \secref{beyond}.  
The chapter concludes in \secref{perspectives} with some personal
views on the use of epistemic logic and epistemic concepts in security
research. 
My observations should not be particularly controversial, but my main
conclusion remains that progress beyond the current state of the art
in security research---at least in security research that benefits
from epistemic logic---will require a deeper understanding of
resource-bounded knowledge, which is itself an active research area in
epistemic logic. 

All bibliographic references are postponed to \secref{bibnotes}, where
full references and additional details are given for 
topics covered in the main body of the chapter.
It is worth noting that the literature on reasoning about security
draws from several fields besides logic. For instance, much
of the research on cryptographic protocols derives from earlier work in
distributed computing. Similarly, recent research both on
cryptographic protocol analysis and on confidentiality in multi-level
systems is based on work in programming language semantics and
static analysis. The interested reader is invited to follow the
references given in \secref{bibnotes} for details.


\section{Cryptographic Protocols}\label{chap13:cryptoprotocols}

Cryptographic protocols are communication protocols---rules for
exchanging messages between agents---that use cryptography to achieve
a security goal such as authenticating one agent to another, or
exchanging confidential messages.  

Cryptographic protocols are a popular object of study for several
reasons. First, they are concrete---they correspond to actual
artifacts implemented and used in practice. 
Second, their theory extends that of distributed protocols
and network protocols in general, which are themselves thoroughly
studied.

Cryptographic protocols have characteristics that distinguish them
from more general communication protocols. In particular, they
\begin{enumerate}

\item[(1)] enforce security properties;

\item[(2)] rely on cryptography;

\item[(3)] execute in the presence of attackers that might 
  attempt to subvert them.

\end{enumerate}

Protocols can be analyzed concretely or symbolically. The concrete
perspective views protocols as
exchanging messages consisting of sequences of bits and subject to
formatting requirements, which is the perspective used in most network
protocols research. 
The symbolic perspective views protocols as exchanging messages
consisting of symbols in some formal language, which is the
perspective used in most distributed protocols research.
The focus of this section is on symbolic cryptographic protocols
analysis.


\subsection{Protocols}
\label{chap13:protocols}

A common notation for protocols is to list the sequence of messages
exchanged between the parties involved in the protocol, since the kinds
of protocols studied rarely involve complex control flow. 

A simple protocol between Alice and
Betty (represented by $A$ and $B$) in which Alice sends
message $m_1$ to Betty and Betty responds by sending message $m_2$ to
Alice would be described by:
\begin{protocol}{sample}
\pmessage{1}{A}{B}{m_1}\\
\pmessage{2}{B}{A}{m_2}
\end{protocol}%
The message sequence notation takes a global view of the protocol, 
describing the protocol from the outside, so to speak. 
An alternate way to describe a protocol is to specify the
roles of the parties involved in the protocol. For
protocol~\eqref{sample}, for instance, there are two roles: the
initiator role, who sends message $m_1$ to the receiver and waits
for a response message, and the receiver role, who waits for a
message to arrive from the initiator and responds with $m_2$.

A protocol executes in an environment, which details anything relevant
to the execution of said protocol, such as the agents participating in
the protocol, whether other instances of the protocol are also
executed concurrently, the possible attackers and their capabilities.
The result of executing a protocol in a given
environment can be modeled by a set of traces, where a trace
corresponds to a possible execution of the protocol.
A trace is a sequence of global states. 
A global state records the local state of every agent involved in the
protocol, as well as the state of the environment.  
This general description is compatible with most representations in
the literature, and can be viewed as a Kripke structure by defining a
suitable accessibility relation over the states of the system.

To illustrate protocols in general, and initiate the study of
confidentiality, here are two simple protocols that achieve a specific
form of confidentiality without requiring cryptography. 
One lesson to be drawn from these examples is that confidentiality in
some cases can be achieved without complex operations. 

The first protocol solves an instance of the following problem: how
two agents may exchange secret information in the open, without an 
eavesdropping third agent learning about the information.
The instance of the problem, called the Russian Cards problem, is
pleasantly concrete and can be explained to children: Alice and Betty
each draw three cards from a pack of seven cards, and Eve (the
eavesdropper) gets the remaining card. 
Can players Alice and Betty learn each other's cards without
revealing that information to Eve? The restriction is that Alice
and Betty can only make public announcements that Eve can hear.

Several protocols for solving the Russian Cards problem have been
proposed; a fairly simple solution is the \emph{Seven Hands protocol}. 
Recall that there are seven cards: three are dealt to Alice, three are
dealt to Betty, and the last card is dealt to Eve. Call the cards
dealt to Alice $a_1,a_2,a_3$, and the cards dealt to Betty 
$b_1,b_2,b_3$. The card dealt to Eve is $e$. 

The Seven Hands protocol is a two-step protocol that Alice can use to
tell her cards to Betty and learn Betty's cards in response:
\begin{protocol}{seven-hands}
\pmessage{1}{A}{B}{\mi{SH}_A}\\
\pmessage{2}{B}{A}{\mi{SH}_B}
\end{protocol}%
where $\mi{SH}_A$ and $\mi{SH}_B$ are the following specific messages:
\begin{enumerate}
\item[(1)] Message $\mi{SH}_A$ is constructed by
Alice as follows. Alice first chooses a random renaming $W,X,Y,Z$ of
the elements in $\{b_1,b_2,b_3,e\}$, that is, a random permutation of
the four cards not in her hand. Message $\mi{SH}_A$ then consists of
the following seven subsets of cards, in some arbitrary order:
\begin{gather*}
   \{a_1,a_2,a_3\}\\
   \{a_1,W,X\}\quad\{a_1,Y,Z\}\\
   \{a_2,W,Y\}\quad\{a_2,X,Z\}\\
   \{a_3,W,Z\}\quad\{a_3,X,Y\}
\end{gather*}
These subsets are carefully chosen: for every possible hand of Betty,
that is, for every possible subset $S$ of size three of $\{W,X,Y,Z\}$,
there is exactly one set in $\mi{SH}_A$ with which $S$ has an empty
intersection, and that set is Alice's hand $\{a_1,a_2,a_3\}$. 
Thus, upon receiving $\mi{SH}_A$, Betty can identify Alice's hand by
examining the sets Alice sent and picking the one with which her own
hand has an empty intersection, and in the process Betty learns
Alice's hand, and by elimination, Eve's card. 

\item[(2)] Message $\mi{SH}_B$, Betty's response, is simply Eve's card. 
Alice, upon receiving $\mi{SH}_B$, knows her own hand and Eve's card,
and therefore can infer by elimination Betty's hand. 
\end{enumerate}
At the end of the exchange, Alice knows Betty's hand and Betty knows
Alice's hand, as required.

What about Eve? She does not learn anything about the cards in Alice
or Betty's hand. Indeed, after seeing Alice's message, Eve has no
information about Alice's 
hand, since every card appears in exactly three of the sets Alice
sent. There is no way for Eve to isolate which of those cards might
be one of Alice's. Furthermore, after seeing Betty's message, all she has
learned is her own card, which she already knew.

If we define $c\cin i$ to be the primitive proposition \emph{card $c$
  is in player $i$'s hand} (where $A,B,E$ represent Alice, Betty, and
Eve, respectively) then we expect that the following
epistemic formula holds after the first message is received by Betty:
\[K_B(a_1\cin A) \land K_B(a_2\cin A) \land K_B(a_3\cin A),\]
that the following epistemic formula holds after the second message is
received by Alice:
\[ K_A(b_1\cin B) \land K_A(b_2\cin B) \land K_B(b_3\cin B),\]
and
that the following formula holds after either of the messages is received:
\begin{align*}
& \neg K_E(a_1\cin A) \land \neg K_E(a_2\cin A) \land \neg K_E(a_3\cin
A)\\
& \qquad \land \neg K_E(b_1\cin B)\land \neg K_E(b_2\cin B)\land \neg K_E(b_3\cin B).
\end{align*} 
It is an easy exercise to construct the
Kripke structures describing this scenario.

The Seven Hands protocol is ideally suited for epistemic
reasoning via a possible-worlds semantics for knowledge, as it relies
on combinatorial analysis. Its applicability, however, is limited. 

The second protocol is a protocol to ensure anonymity, which is a form
of confidentiality (see \secref{beyond}). 
It does not rely on combinatorial analysis but rather on 
properties of the XOR operation.\footnote{XOR (exclusive or) is a
  binary Boolean operation $\oplus$ defined by taking $b_1 \oplus b_2$
  to be true if and only if exactly one of $b_1$ or $b_2$ is true. It
  is associative and commutative.}
The Dining Cryptographers protocol was originally developed to solve
the following problem. Suppose that Alice, Betty, and Charlene are three
cryptographers having dinner at their favorite restaurant.
Their waiter informs them that arrangements have been made for the
bill to be paid anonymously by one party.
That payer might be one of the cryptographers, but it might also be
U.S. National Security Agency. The three cryptographers 
respect each other's right to make an anonymous payment, but they
would like to know whether the NSA is paying.

The following protocol can be used to satisfy the cryptographers'
curiosity and allow each of them to determine whether the NSA or one
of her colleagues is paying, without revealing 
the identity of the payer in the latter case. 
\begin{enumerate}

\item[(1)] Every cryptographer $i$ flips a fair coin privately with her
  neighbor $j$ on her right: the Boolean result $T_{\{i,j\}}$ is 
$\mi{true}$ if the coin lands tails, and $\mi{false}$ if
  the coin lands heads. Thus, the
  cryptographers produce the Boolean results $T_{\{A,B\}}$,
  $T_{\{A,C\}}$, $T_{\{B,C\}}$; Alice sees $T_{\{A,B\}}$ and
  $T_{\{A,C\}}$; Betty sees $T_{\{A,B\}}$ and $T_{\{B,C\}}$; Charlene sees
  $T_{\{A,C\}}$ and $T_{\{B,C\}}$. 

\item[(2)] Every cryptographer $i$ computes a private Boolean value $\mi{Df}_i$ as
  $\mi{true}$ if the two coin tosses she has witnessed are
  different, and $\mi{false}$ if they are the same. Thus, $\mi{Df}_A =
  T_{\{A,B\}}\oplus T_{\{A,C\}}$, $\mi{Df}_B = T_{\{A,B\}}\oplus
  T_{\{B,C\}}$, and $\mi{Df}_C = T_{\{A,C\}}\oplus T_{\{B,C\}}$. 

\item[(3)] Every cryptographer $i$ publicly announces 
  $\mi{Df}_i$, except for the paying cryptographer (if there is one) who announces $\neg
  \mi{Df}_i$, the negation of $\mi{Df}_i$.  

\end{enumerate}
Once the protocol is executed, any curious cryptographer interested in
determining who paid for dinner simply has to take the 
XOR of all the announcements: if the result is $\mi{false}$, then the NSA
paid, and if the result is $\mi{true}$, then one of the 
cryptographers paid.

To see why this is the case, consider the two possible scenarios.
Suppose the NSA paid. Then the XOR of all
the announcements is:
\begin{align*}
 & \mi{Df}_A \oplus \mi{Df}_B \oplus \mi{Df}_C \\
 & \qquad = \left(T_{\{A,B\}}\oplus T_{\{A,C\}}\right) \oplus
   \left(T_{\{B,C\}}\oplus T_{\{A,B\}}\right) \oplus
   \left(T_{\{A,C\}}\oplus T_{\{B,C\}}\right) \\
 & \qquad = 
   \left(T_{\{A,B\}}\oplus T_{\{A,B\}}\right) \oplus
   \left(T_{\{B,C\}}\oplus T_{\{B,C\}}\right) \oplus
   \left(T_{\{A,C\}}\oplus T_{\{A,C\}}\right) \\
 & \qquad = \mi{false} \oplus \mi{false} \oplus \mi{false} \\
 & \qquad = \mi{false}
\end{align*}
whereas if one of the cryptographers paid (without loss of generality,
suppose it is Alice), then the XOR of all the announcements is:
\begin{align*}
 & \neg \mi{Df}_A \oplus \mi{Df}_B \oplus \mi{Df}_C \\
 & \qquad = 
   \neg \left(T_{\{A,B\}}\oplus T_{\{A,C\}}\right) \oplus
   \left(T_{\{B,C\}}\oplus T_{\{A,B\}}\right) \oplus
   \left(T_{\{A,C\}}\oplus T_{\{B,C\}}\right) \\
 & \qquad = 
   \left(\neg T_{\{A,B\}}\oplus T_{\{A,C\}}\right) \oplus
   \left(T_{\{B,C\}}\oplus T_{\{A,B\}}\right) \oplus
   \left(T_{\{A,C\}}\oplus T_{\{B,C\}}\right) \\
 & \qquad = 
   \left(\neg T_{\{A,B\}}\oplus T_{\{A,B\}}\right) \oplus
   \left(T_{\{B,C\}}\oplus T_{\{B,C\}}\right) \oplus
   \left(T_{\{A,C\}}\oplus T_{\{A,C\}}\right) \\
 & \qquad = \mi{true} \oplus \mi{false} \oplus \mi{false} \\
 & \qquad = \mi{true}
\end{align*}
If one of the cryptographers paid, neither of the two other
cryptographers will know which of her colleagues paid, since either
possibility is compatible with what they can observe. 
Again, it is an easy exercise to construct the Kripke
structures capturing these scenarios.


\subsection{Cryptography}
\label{chap13:crypto}

While protocols such as the Seven Hands protocol and the Dining
Cryptographers protocol enforce confidentiality by carefully
constructing specific messages meant to convey specific information in
a specific context, most cryptographic protocols rely on cryptography
for confidentiality. 

Cryptography seems a natural approach for confidentiality.  
After all, the whole point of cryptography is to hide information in
such a way that only agents with a suitable key can access the
information.
And indeed, if the goal is for Alice to send message $m$ to Betty
when Alice and Betty alone share a key to encrypt and decrypt messages,
then the simplest protocol for confidential message exchange 
is simply for Alice to encrypt $m$ and send it to Betty. 
But how do Alice and Betty come to share a key in the first place?
Distributing keys is tricky, because keys have to be sent to the right
agents, in such a way that no other agent can get them. 

Before addressing those problems, let us review the basics of
cryptography. The reader is assumed to have been exposed to at least
informal descriptions of cryptography.
An encryption scheme is defined by a set of sourcetexts, a
set of ciphertexts, a set of keys, and for every key $k$ an injective
encryption function $e_k$ producing a ciphertext from a sourcetext and
a decryption function $d_k$ producing a sourcetext from a ciphertext, 
with the property that $d_k(e_k(x))=x$ for all sourcetexts
$x$. We often assume that ciphertexts and keys are included in
sourcetexts to allow for nested encryption and encrypted
keys.\footnote{This section considers deterministic  
  encryption schemes only, ignoring probabilistic encryption schemes.}

There are two broad classes of encryption schemes, which differ in
how keys are used for decryption. \emph{Shared-key encryption schemes}
require an agent to have a full key to both encrypt and decrypt a
message. 
They tend to be efficient, and can often be implemented directly in
hardware. \emph{Public-key encryption schemes},
on the other hand, are set up so that an
agent only needs to know part of a key (called the public key)
to encrypt a message, while needing the full key to decrypt a
message. The full key cannot be easily recovered from knowing only the
public key. 
Public keys are generally made public (hence the
name), so that any agent can encrypt a message intended for, say,
Alice, by looking up and using Alice's public key. Since
only Alice has the full key, only she can decrypt that message. 
DES and AES are concrete examples of shared-key encryption schemes,
while RSA and elliptic-curve encryption schemes are concrete examples 
of public-key encryption schemes.

Cryptographic protocols are needed with shared-key encryption schemes
because agents need to share a key in order to exchange encrypted
messages.
How is such a shared key distributed? 
And how can agents make sure they are not tricked into sharing those
keys with attackers?
An additional difficulty is that when the same shared key is reused
for every interaction between two agents, the content of all those
interactions becomes available to an attacker that manages to learn
that key. 
To minimize the impact of a key compromise, many systems create a
fresh session key for any two agents that want to
communicate, which exacerbates the key distribution problem. 

Public-key encryption simplifies key distribution, since public keys
can simply be published. Any agent wanting to send a confidential
message to Alice has only to look up Alice's public key and use it to
encrypt her message. The problem, from Alice's perspective, is that
anyone can encrypt a message and send it to her, which means that if
Alice wants to make sure that the encrypted message she received came
from Betty, some sort of authentication mechanism is needed. 
Furthermore, all known public-key encryption schemes are
computationally expensive, so a common approach is to have agents that
want to exchange messages in a session first use public-key encryption
to generate a session key for a shared-key encryption scheme that they
use for their exchanged messages. Such a scenario requires
authentication to ensure that agents are not tricked into
communicating with an attacker.


\paragraph{Sample Cryptographic Protocols.}

Most classical cryptographic protocols are designed to solve the
problem of key distribution for shared-key encryption schemes, and of
authentication for public-key encryption schemes. 
In these contexts, confidentiality and authentication are the key
properties: confidentiality to enforce that distributed keys remain
secret from attackers, and authentication to ensure that agents can
establish the identity of the other agents involved in a message
exchange.  

This section presents two protocols, each illustrating different
problems that can arise and highlighting vulnerabilities that
attackers can exploit. 
(Attackers will be introduced more carefully in the next section.)
The first protocol distributes session keys for a shared-key
encryption scheme, while the second protocol aims at achieving mutual
authentication for public-key encryption schemes.
Not all of the problems illustrated will occur in every protocol, of
course, nor are vulnerabilities in one context necessarily
vulnerabilities in another context. 

For the first protocol, consider the following situation. Suppose
Alice wants to communicate with Betty, and there is a trusted server
Serena who will generate a shared session key (for some shared-key
encryption scheme) for them to use. Assume that every registered user
of the system shares a distinct key with the trusted server in some
shared-key encryption scheme; these keys for Alice and Betty are
denoted $k_{\mi{AS}}$ and $k_{\mi{BS}}$.

The idea is for Serena to generate a fresh key and send it
to both Alice and Betty. 
Sending it in the clear, however, would allow an eavesdropping
attacker to 
read
it and then use it to decrypt messages between Alice and Betty.
Since Alice and Betty both share a key with Serena, one solution might
be to use those keys to encrypt the session key sent to Alice and 
Betty, but this turns out to be difficult to implement in practice. 
Here is the problem. Alice, wanting to communicate with Betty, sends a
message to Serena asking her to generate a session key, and Serena
sends it to both Alice and Betty. As far as Betty is concerned, she
receives a key with an indication that Alice will use it to send her
messages. Betty now has to store the key and wait for Alice to send
her messages encrypted with that key. If Alice wants to set up several
concurrent communications with Betty, then Betty will have to match
each incoming communication with the appropriate key, which is
annoying at best and inefficient at worst. It turns out to be more
efficient for Serena to send the fresh session key $k_{\mi{sess}}$ to
Alice, and for Alice to forward the key to Betty in her first
message. This observation leads to the following protocol:
\begin{protocol}{prot1}
\pmessage{1}{A}{S}{B}\\
\pmessage{2}{S}{A}{B,\sencr{k_{\mi{sess}}}{k_{\mi{AS}}},\sencr{k_{\mi{sess}}}{k_{\mi{BS}}}}\\
\pmessage{3}{A}{B}{A,\sencr{k_{\mi{sess}}}{k_{\mi{BS}}}}
\end{protocol}%
Both Alice and Betty learn key $k_{\mi{sess}}$, which is kept secret from eavesdroppers.

While this protocol might seem sufficient to distribute a key to both
Alice and Betty, several things can go wrong in the presence of an
insider attacker, that is, an attacker that is also a registered user
of the system and has control over the network (i.e., can intercept
and forge messages; see \secref{attackers}). 

The insider attacker, Isabel, can initiate a communication with
trusted server Serena via her shared key $k_{\mi{IS}}$ (Isabel is
assumed to have such a key because she is a registered user of the
system), and use the key to pose as Alice to Betty. Here is a
sequence of messages exemplifying the attack, where the notation
$I[A]$ denotes $I$ posing as $A$:\footnote{We can assume that
every message has a \emph{from} and \emph{to} field---think
email---and that these can be forged. Isabel posing as Alice means
that Isabel sends a message and forges the \emph{from} field of the
message to hold Alice's name.}
\begin{interaction}{attack2}
\imessage{I}{S}{B}\\
\imessage{S}{I}{B,\sencr{k_{\mi{sess}}}{k_{\mi{IS}}},\sencr{k_{\mi{sess}}}{k_{\mi{BS}}}}\\
\imessage{I[A]}{B}{A,\sencr{k_{\mi{sess}}}{k_{\mi{BS}}}}
\end{interaction}%
Betty believes that she is sharing key $k_{\mi{sess}}$ with Alice,
while she is in fact sharing it with Isabel. 
This is a failure of authentication---the protocol does not 
authenticate the initiator to the responder.

Isabel can also trick Alice into believing she is talking to Betty, by
posing as the server and intercepting messages between Alice and the
server, as the following sequence of messages exemplifies: 
\begin{interaction}{attack3}
\imessage{A}{I[S]}{B}\\
\imessage{I[A]}{S}{I}\\
\imessage{S}{I[A]}{I,\sencr{k_{\mi{sess}}}{k_{\mi{AS}}},\sencr{k_{\mi{sess}}}{k_{\mi{IS}}}}\\
\imessage{I[S]}{A}{B,\sencr{k_{\mi{sess}}}{k_{\mi{AS}}},\sencr{k_{\mi{sess}}}{k_{\mi{IS}}}}\\
\imessage{A}{I[B]}{A,\sencr{k_{\mi{sess}}}{k_{\mi{IS}}}}
\end{interaction}%
This form of attack is commonly known as a man-in-the-middle 
attack. 
Isabel intercepts Alice's message to the server, and turns around and
sends a different message to the server posing as Alice. 
The response from the server
is intercepted by Isabel, who crafts a suitable response back to
Alice. Alice takes that response (which she believes is coming from
the server) and sends it to Betty, but that message is intercepted by
Isabel as well. Now, as far as Alice is concerned, she has successfully
completed the protocol, and holds a key $k_{\mi{IS}}$ that she believes
she can use to communicate confidentially with Betty, while she is
really communicating with Isabel. 

How can we correct these vulnerabilities? 
One feature on which these attacks rely is that the identity of the
intended parties for the keys in the protocol are easily forged by the
attacker. So one fix is to bind the intended parties to the
appropriate copies of the key. Here is an amended version of the protocol:
\begin{protocol}{prot4}
\pmessage{1}{A}{S}{B}\\
\pmessage{2}{S}{A}{B,\sencr{B,k_{\mi{sess}}}{k_{\mi{AS}}},\sencr{A,k_{\mi{sess}}}{k_{\mi{BS}}}}\\
\pmessage{3}{A}{B}{\sencr{A,k_{\mi{sess}}}{k_{\mi{BS}}}}
\end{protocol}%
When Alice receives her response from the server and decrypts her
message $\sencr{B,k_{\mi{sess}}}{k_{\mi{AS}}}$, she can verify that the key she meant
Serena to create to communicate with Betty is in fact a key meant to
communicate with Betty. This suffices to foil Isabel in
attack~\eqref{attack3}. 
Similarly, when Betty receives her message from Alice containing
$\sencr{A,k_{\mi{sess}}}{k_{\mi{BS}}}$, she can verify that the key is meant to
communicate with Alice. This suffices to foil attack~\eqref{attack2}.

Protocol~\eqref{prot4} now seems to work as intended. 
It does suffer from another potential vulnerability, though, one that
is less directly threatening, but can still cause problems: it is
susceptible to a replay attack.
Here is the scenario.
Suppose that Isabel eavesdrops on messages as Alice gets a session key
$k_0$ from the trusted server to communicate with Betty, and holds on
to messages $\sencr{B,k_0}{k_{\mi{AS}}}$ and
$\sencr{A,k_0}{k_{\mi{BS}}}$. Suppose further that after a long delay
Isabel manages to somehow obtain key $k_0$, perhaps by breaking into
Alice's or Betty's computer, or by expending several months' worth of
effort to crack the encryption.
Once Isabel has $k_0$, she can subvert an attempt by Alice to get a
session key for communicating with Betty by simply intercepting the
messages from Alice to Serena, and replaying the messages
$\sencr{B,k_0}{k_{\mi{AS}}}$ and $\sencr{A,k_0}{k_{\mi{BS}}}$ she
intercepted in the past. The following sequence of messages
exemplifies this attack:
\begin{interaction}{attack5}
\imessage{A}{I[S]}{B}\\
\imessage{I[S]}{A}{B,\sencr{B,k_0}{k_{\mi{AS}}},\sencr{A,k_0}{k_{\mi{BS}}}}\\ 
\imessage{A}{B}{\sencr{A,k_0}{k_{\mi{BS}}}}
\end{interaction}%
The main point here is that Alice and Betty after this protocol
interaction end up using key $k_0$ as their session key, but that key
is one that Isabel knows, meaning that Isabel can decrypt every single
message that Alice and Betty exchange in that session. 
So even though she does not have the shared keys $k_{\mi{AS}}$ and
$k_{\mi{BS}}$, she has managed to trick Alice and Betty into using a
key she knows.

Preventing this kind of replay attack requires ensuring that messages
from earlier executions of the protocol cannot be used in later
executions. 
One way to do that is to have every agent record every message they
have ever sent and received, but that is too expensive to be
practical. 
The common alternative is to use timestamps, or nonces. A nonce is a
large random number, meant to be unpredictable and essentially
unique---the likelihood that the same nonce occurs twice
within two different sessions should be negligible. 
To fix protocol~\eqref{prot4} and prevent replay attacks, it suffices
for Alice and Betty to generate nonces $n_A$ and $n_B$, respectively,
and send them to trusted server Serena so that she can include them in
her responses:
\begin{protocol}{prot6}
\pmessage{1}{A}{B}{n_A}\\
\pmessage{2}{B}{S}{A,n_A,n_B}\\
\pmessage{3}{S}{A}{\sencr{B,k_{\mi{sess}},n_A}{k_{\mi{AS}}},\sencr{A,k_{\mi{sess}},n_B}{k_{\mi{BS}}}}\\
\pmessage{4}{A}{B}{\sencr{A,k_{\mi{sess}},n_B}{k_{\mi{BS}}}}
\end{protocol}%
As long as Alice and Betty, when each receives her encrypted message
containing the session key, both check that the nonce in the encrypted
message is
the one that they generated, then they can be confident that the
encrypted messages have not been reused from earlier sessions. 

Protocol~\eqref{prot6} now seems to work as intended and is not
vulnerable to replay attacks. 
But it does not actually guarantee mutual authentication; that is, it
does not guarantee to Alice that she is in fact talking to Betty when
she believes she is, and to Betty that she is in fact talking to Alice
when she believes she is. Consider the following attack, in which
attacker Trudy is not an insider---she is not a registered user of the
system---but has control of the network and thus can intercept and
forge messages. Trudy poses as Betty by intercepting messages from
Alice and forging responses:
\begin{interaction}{attack7}
\imessage{A}{T[B]}{n_A}\\
\imessage{T[B]}{S}{A,n_A,n_T}\\
\imessage{S}{A}{\sencr{B,k_{\mi{sess}},n_A}{k_{\mi{AS}}},\sencr{A,k_{\mi{sess}},n_T}{k_{\mi{BS}}}}\\
\imessage{A}{T[B]}{\sencr{A,k_{\mi{sess}},n_T}{k_{\mi{BS}}}}
\end{interaction}%
From Alice's perspective, she has completed the protocol by exchanging
messages with Betty, and holds a session key for sending confidential
messages to Betty. But of course Alice has been communicating with
Trudy, and Betty is not even aware of the exchange. Trudy cannot
actually read the messages sent by Alice, so there is no breach of
confidentiality, but Trudy has managed to trick Alice into believing
she shares a session key with Betty. 
In terms of knowledge, Alice knows the session key, but she does not
know that Betty does.

There is also a way for Trudy to trick Betty into believing she shares
a session key with Alice, by posing as Alice:
\begin{interaction}{attack8}
\imessage{T[A]}{B}{n_T}\\
\imessage{B}{S}{A,n_T,n_B}\\
\imessage{S}{T[A]}{\sencr{B,k_{\mi{sess}},n_T}{k_{\mi{AS}}},\sencr{A,k_{\mi{sess}},n_B}{k_{\mi{BS}}}}\\
\imessage{T[A]}{B}{\sencr{A,k_{\mi{sess}},n_B}{k_{\mi{BS}}}}
\end{interaction}%
From Betty's perspective, she has completed the protocol by exchanging
messages with Alice, and holds a session key for sending confidential
messages to Alice. But of course Betty has been communicating with
Trudy, and Alice is not even aware of the exchange. 
In terms of knowledge, Betty knows the
session key, but she does not know that Alice does.

Mutual authentication is achieved through an additional nonce
exchange at the end of the protocol which uses the newly created
session key:
\begin{protocol}{prot9}
\pmessage{1}{A}{B}{n_A}\\
\pmessage{2}{B}{S}{A,n_A,n_B,n'_B}\\
\pmessage{3}{S}{A}{n'_B,\sencr{B,k_{\mi{sess}},n_A}{k_{\mi{AS}}},\sencr{A,k_{\mi{sess}},n_B}{k_{\mi{BS}}}}\\
\pmessage{4}{A}{B}{n'_A,\sencr{A,k_{\mi{sess}},n_B}{k_{\mi{BS}}},\sencr{A,n'_B}{k_{\mi{sess}}}}\\
\pmessage{5}{B}{A}{\sencr{B,n'_A}{k_{\mi{sess}}}}
\end{protocol}%

This protocol now seems to work as intended without being vulnerable
to replay attacks or authentication failures. How can this be
guaranteed?

Intuitively, protocol~\eqref{prot9} is not susceptible to replay
attacks because of the use of nonces: Alice can deduce that the first
encrypted component of the third message was not reused from an
earlier protocol execution, and Betty can deduce that the first
encrypted component of the fourth message was not reused from an
earlier protocol execution.

Similarly, mutual authentication in protocol~\eqref{prot9} follows from the
use of shared keys: Alice can deduce that Serena created the first
encrypted component of the third message; and Betty can deduce that
Serena created the first encrypted component of the fourth
message. 
Moreover, if Betty believes that $k_{\mi{sess}}$ is a key known only
to Alice and herself, then she can deduce that Alice created the
second encrypted component in the fourth message, and similarly, if
Alice believes that $k_{\mi{sess}}$ is a key known only to Betty and
herself, then she can deduce that Betty created the
encrypted component in the fifth message.

The confidentiality of the session key requires the assumption that
trusted server Serena is indeed trustworthy and creates keys that
have not previously been used and distributed to other parties. 
If so, then 
Alice can deduce that the session key she receives in the third
message is a confidential key for communicating with Betty, since
Alice can also deduce that she has been executing the protocol with
Betty.
Similarly, Betty can deduce that the session key she receives in the
fourth message is a confidential key for communicating with Alice,
since Betty can deduce that she has been executing the protocol with
Alice

In a precise sense, the goal of cryptographic protocol analysis is to
prove these kind of properties formally, and many techniques have been
developed which are surveyed below in \secref{reasoning}.

The second cryptographic protocol uses a public-key
encryption scheme to achieve mutual authentication:
\begin{protocol}{needham-schroeder}
\pmessage{1}{A}{B}{\sencr{A,n_A}{\mi{pk}_B}}\\
\pmessage{2}{B}{A}{\sencr{n_A,n_B}{\mi{pk}_A}}\\
\pmessage{3}{A}{B}{\sencr{n_B}{\mi{pk}_B}}
\end{protocol}%
where $\mi{pk}_A$ and $\mi{pk}_B$ are the public keys of Alice and
Betty, respectively, and $n_A$ and $n_B$ are nonces.

Intuitively, when Alice receives her nonce $n_A$ back in the second
message, she knows that Betty must have decrypted her first message at
some point during the execution of the protocol (because only Betty
could have decrypted the message that contained it), and similarly,
when Betty receives her nonce $n_B$ back, she knows that Alice must
have decrypted her second message. Note that $n_A$ and $n_B$ are kept
confidential throughout the protocol, and that mutual authentication
relies on that confidentiality.

Protocol~\eqref{needham-schroeder}, known as the Needham-Schroeder
protocol, achieves mutual authentication even in the presence of an
attacker that has control of the network and can intercept and forge
messages. It is however vulnerable to insider
attackers that are registered users of the system and have control of
the network. 
For example, insider attacker Isabel can use an attempt by Alice to
initiate an authentication session with her to trick unsuspecting
Betty into believing that Alice is initiating an authentication
session with her:
\begin{interaction}{attack-ns}
\imessage{A}{I}{\sencr{A,n_A}{\mi{pk}_I}}\\
\imessage{I[A]}{B}{\sencr{A,n_A}{\mi{pk}_B}}\\
\imessage{B}{I[A]}{\sencr{n_A,n_B}{\mi{pk}_A}}\\
\imessage{I}{A}{\sencr{n_A,n_B}{\mi{pk}_A}}\\
\imessage{A}{I}{\sencr{n_B}{\mi{pk}_I}}\\
\imessage{I[A]}{B}{\sencr{n_B}{\mi{pk}_B}}
\end{interaction}%
From Alice's perspective, she has managed to complete a mutual
authentication session with Isabel, which was her goal all
along. 
But Isabel also managed to complete an authentication session with
Betty, tricking Betty into believing she is interacting with Alice.

There is a simple fix that eliminates that vulnerability:
\begin{protocol}{needham-schroeder-lowe}
\pmessage{1}{A}{B}{\sencr{A,n_A}{\mi{pk}_B}}\\
\pmessage{2}{B}{A}{\sencr{B,n_A,n_B}{\mi{pk}_A}}\\
\pmessage{3}{A}{B}{\sencr{n_B}{\mi{pk}_B}}
\end{protocol}%
It is interesting to see how the fix works: if Alice, during her
mutual authentication attempt with Isabel, notices that the response
message she receives from Isabel names a different agent than Isabel, 
then she can deduce that her authentication attempt is being
subverted to try to confound another agent, and she can abort the
authentication attempt at that point.


\subsection{Attackers}\label{chap13:attackers}

A distinguishing feature of cryptographic protocols, besides the use
of cryptography, is that they are deployed in potentially
hostile environments in which attackers
may attempt to subvert the
operations of the protocol.

Reasoning about cryptographic protocols, therefore, requires a threat
model, describing the kind of attackers against which the
cryptographic protocol should protect. Attackers commonly considered
in the literature include:
\begin{itemize}

\item \textbf{Eavesdropping attackers:} assumed to be able to read all
  messages exchanged between 
  agents. Eavesdropping 
  attackers do not affect communication in any way, however, and
  remain hidden from other agents.

\item \textbf{Active attackers:} assumed to have complete control over
  communications between agents, that is, able to read all messages as
  well as intercept them and forge new messages. They remain hidden
  from other agents, and thus no agent will intentionally attempt to
  communicate with an active attacker.

\item \textbf{Insider attackers:} assumed to have complete control
  over communications between agents just like active attackers, but
  also considered legitimate registered users in their own right. They can
  therefore initiate interactions with other agents as themselves, and
  other agents can intentionally initiate interactions with them.\footnote{A
  fourth class of attackers, less commonly considered, shares 
characteristics with both eavesdropping attackers and insider attackers:
dishonest agents are assumed not to have control over the network but
may attempt to subvert the protocol while acting within the limits
imposed on legitimate users.}

\end{itemize}
The class of insider attackers includes the class of active attackers,
which itself includes the class of eavesdropping attackers.  
Thus, in that sense, an insider attacker is stronger than an active
attacker which is stronger than an eavesdropping attacker. In
practice, this means that a cryptographic protocol that is deemed
secure in the presence of an insider attacker will remain so in the
presence of active and eavesdropping attackers, and so on.

We saw several examples of attacks in \secref{crypto}, performed by
different kind of attackers. Most of the protocols in \secref{crypto}
achieve their goals in the presence of eavesdropping attackers, while some
also achieve their goals in the presence of active attackers but fail
in the presence of insider attackers. 
The Needham-Schroeder protocol~\eqref{needham-schroeder}, for
instance, can be shown to satisfy 
mutual authentication in the presence of active attackers, but not in
the presence of insider attackers---as exemplified by
attack~\eqref{attack-ns}---while the variant
protocol~\eqref{needham-schroeder-lowe} achieves mutual 
authentication even in the presence of insider attackers.

The attacks described in \secref{crypto} took place at the level of
the protocols themselves, and not at the level of the encryption
schemes used by the protocols. 
But vulnerabilities in encryption schemes are also relevant: an
attacker cracking an encrypted message from the trusted server to the
agents in protocol~\eqref{prot9} will learn the session key, which will
invalidate any confidentiality guarantees claimed for the protocol. 
Despite this, cryptographic protocol are typically analyzed
independently from the details of the encryption scheme. 
The main reason is that it abstracts away from vulnerabilities specific
to the encryption scheme used, leaving only those relating to the
cryptographic protocol.  
Vulnerabilities in encryption schemes are usually independent of the
cryptographic protocols that use them, and can be investigated
separately.
A vulnerability in the protocol will be a
vulnerability no matter what 
encryption scheme is used, and requires a change in the protocol to
correct the flaw. 

The standard way to analyze cryptographic protocols independently of
any encryption scheme is to use a \emph{formal model of cryptography}
that assumes perfect encryption leaking no information about
encrypted content. It can be defined as the following symbolic
encryption scheme.
If $P$ is a set of plaintexts and $K$ is a set of keys, then the
set of sourcetexts is taken to be the smallest set $S$ of symbolic
terms containing $P$ and
$K$ such that $(x,y)\in S$ and $\sencr{x}{k}\in S$ when $x,y\in S$ and
$k\in K$.  
Intuitively, $(x,y)$ represents the concatenation of $x$ and $y$, and
$\sencr{x}{k}$ represents the encryption of $x$ with key
$k$. The ciphertexts are all sourcetexts of the form
$\sencr{x}{k}$.
The symbolic encryption function $e_k(x)$ simply returns
$\sencr{x}{k}$, and the symbolic decryption function $d_k(x)$
returns $y$ if $x$ is $\sencr{y}{k}$, and some special token
\textbf{fail} otherwise.\footnote{The symbolic decryption function
embodies an assumption that encrypted messages have enough redundancy
for an agent to determine when decryption is successful.}

In the context of analyzing protocols with a formal model of
cryptography, attackers are usually modeled using \emph{Dolev-Yao
  capabilities}. 
These capabilities go hand in hand with the symbolic aspect of formal
models of cryptography. 
Intuitively, eavesdropping Dolev-Yao attackers can split up
concatenated messages
and decrypt them if they know the decryption key; active Dolev-Yao
attackers can additionally create new messages by concatenating
existing messages and encrypting them with known keys. 
Dolev-Yao attackers do not have the capability of cracking
encryptions, nor can they access messages at the level of their
component bits.


\subsection{Modeling Knowledge}\label{chap13:knowledge}

The analyses in \secref{crypto} show that various notions of knowledge
arise rather naturally when reasoning informally about properties of
cryptographic protocols. 
There are essentially two main kinds of knowledge described in the
literature. In some frameworks, both kinds of knowledge are used.

\paragraph{Message Knowledge.}
The first kind of knowledge, the most common and in some sense the
most straightforward, tries to capture the notion of \emph{knowing a
  message}. 

There are several equivalent approaches to modeling this kind of
knowledge, at least in a formal model of cryptography with Dolev-Yao
capabilities. 
Intuitively, the idea is a constructive one: an attacker knows a
message if she can construct that message from other messages she has
received or intercepted. 
(Message knowledge in the context of confidentiality properties is
often presented from the perspective of an attacker, since
confidentiality is breached when the attacker comes to know a particular
message.) 
In such a context, knowing a message is sometimes called 
\emph{having a message}, \emph{possessing a message}, or \emph{seeing
  a message}. 

Message knowledge may be described via the following sets. 
Let $H$ be a set of messages that the attacker has received or
intercepted. 
The set $\mi{Parts}(H)$, the set of all components of messages from
$H$, is defined inductively by the following inference rules:
\begin{gather*}
 \Rule{m \in H}{m\in\mi{Parts}(H)}\qquad
 \Rule{\sencr{m}{k}\in\mi{Parts}(H)}{m\in\mi{Parts}(H)}\\\displaybreak[0]
 \Rule{(m_1,m_2)\in\mi{Parts}(H)}{m_1\in\mi{Parts}(H)}\qquad
 \Rule{(m_1,m_2)\in\mi{Parts}(H)}{m_2\in\mi{Parts}(H)}
\end{gather*}
We see that the content of all encrypted messages in $H$ is included
in $\mi{Parts}(H)$, even those that the attacker cannot
decrypt. In a sense, $\mi{Parts}(H)$ is an upper bound on messages the
attacker can know. 
In contrast, the set $\mi{Analyzed}(H)$ of messages that the attacker
can actually see is more restricted:
\begin{gather*}
 \Rule{m \in H}{m\in\mi{Analyzed}(H)}\\\displaybreak[0]
 \Rule{(m_1,m_2)\in\mi{Analyzed}(H)}{m_1\in\mi{Analyzed}(H)}\qquad
 \Rule{(m_1,m_2)\in\mi{Analyzed}(H)}{m_2\in\mi{Analyzed}(H)}\\\displaybreak[0]
 \Rule{\sencr{m}{k}\in\mi{Analyzed}(H)\quad k\in\mi{Analyzed}(H)}{m\in\mi{Analyzed}(H)}
\end{gather*}
Clearly, $\mi{Analyzed}(H)\subseteq\mi{Parts}(H)$. 
One definition of message knowledge is to say that an attacker knows
message $m$ in a state where she has received or intercepted a set $H$
of messages if $m\in\mi{Analyzed}(H)$. 
This is the \emph{attacker
  knows what she can see} interpretation of message knowledge.

The best way to understand this concept of knowledge is to use
a physical analogy: we can think of
plaintext messages as stones, and encrypted messages as locked
boxes. Encrypting a message means putting it in a box and locking it. 
A message is known if it can be held in one's hands. An encrypted
message is known because the box can be held. The content of an
encrypted message is known only if the box can be opened (decrypted)
and the content (a stone or another box) taken and held.

This form of message knowledge can be captured fairly easily in any
logic without using heavy technical machinery, since the data required to
define message knowledge is purely local. If we let
$\mi{Messages}_i(s)$ be the set of messages received or intercepted by
agent $i$ in state $s$
of the system, then we can capture message knowledge via a 
proposition $\mi{knows}_i(m)$, where $i$ is an agent and $m$ is a
message, defined to be true at state $s$ if and only if 
$m\in\mi{Analyzed}(\mi{Messages}_i(s))$. 

Rather than using a dedicated proposition, another approach
relies on a dedicated modal operator to capture message knowledge. 
Message knowledge as defined above can be seen as a form of
\emph{explicit knowledge}, often represented by a modal operator
$X_i\phi$, read \emph{agent $i$ explicitly knows $\phi$}. 
(Explicit knowledge is to be contrasted with the implicit knowledge
captured by the possible-worlds definition of knowledge.)
One form of explicit knowledge, \emph{algorithmic knowledge}, 
uses a local algorithm stored in the local state of an agent to
determine if $\phi$ is explicitly known to that agent. Thus, $X_i\phi$
is true at a state $s$ if the local algorithm of agent $i$ says that
the agent knows $\phi$ in state $s$. 
If we let proposition $\mi{part}_i(m)$ be true at
a state $s$ when $m\in\mi{Parts}(\mi{Messages}_i(s))$, then it is a
simple matter to define a local algorithm to 
check if $m\in\mi{Analyzed}(\mi{Messages}_i(s))$ and capture 
knowledge of message $m$ via $X_i(\mi{part}_i(m))$: \emph{agent $i$
  explicitly knows that message $m$ is part of the messages she has
  received}. 
Thus, $\mi{part}_i(m)$ may be true at a state while
$X_i(\mi{part}_i(m))$ is false at that state if the message is
encrypted with a key that agent $i$ does not know.  

A variant of the \emph{can see} interpretation of
message knowledge is to consider instead the messages that an attacker can
create.
The set $\mi{Synthesized}(H)$ of messages that the attacker can
create from a set $H$ of messages is inductively defined by the
following inference rules:
\begin{gather*}
  \Rule{m \in H}{m\in\mi{Synthesized}(H)}\\\displaybreak[0]
 \Rule{m_1\in\mi{Synthesized}(H)\quad m_2\in\mi{Synthesized}(H)}
        {(m_1,m_2)\in\mi{Synthesized}(H)}\\\displaybreak[0]
  \Rule{m\in\mi{Synthesized}(H)\quad k\in\mi{Synthesized}(H)}
        {\sencr{m}{k}\in\mi{Synthesized}(H)}
\end{gather*}
An alternative interpretation of message knowledge, the \emph{attacker
  knows what she can send} interpretation, can be defined as: an attacker
knows message $m$ in a state where she has received or intercepted a
set $H$ of messages if
$m\in\mi{Synthesized}(\mi{Analyzed}(H))$. Since
$\mi{Analyzed}(H)\subseteq\mi{Synthesized}(\mi{Analyzed}(H))$,
everything an attacker can see she can also send. 

The \emph{can send} interpretation of message knowledge is tricky,
because clearly any agent can send any plaintext and any key---this is
akin to being able to send any password---and it is easy to
inadvertently define nondeterministic attackers that can synthesize
any message. The intent is for attackers to be able to send only
messages based on those she has received or intercepted, but that is a
restriction that can be difficult to justify. This suggests some
subtleties in choosing the right definition of message knowledge. 

Message knowledge, whether under the \emph{can see} or
\emph{can send} interpretation, is severely constrained. It is knowledge
of terms, as opposed to knowledge of facts---although terms can be
facts, facts are more general than terms.
Message knowledge is conducive to formal verification using a
variety of techniques, mostly because it does not require anything but
looking at the local state of an agent. 
Indeed, message knowledge is inherently local.

\paragraph{Possible-Worlds Knowledge.}
The other kind of knowledge that arises in the study of cryptographic
protocols is the standard possible-worlds definition of 
knowledge via an accessibility relation over the states of a structure.
The Kripke structures interpreting knowledge are usually sets of
traces of the protocol
and the
accessibility relation for agent $i$ is an equivalence relation over
the states of the system that relates two states in which agent $i$
has the same local state (including having received or
intercepted the same messages).

In the presence of
cryptography, the standard accessibility relation, meant to capture
when two states are indistinguishable to an agent, seems
inappropriate. 
After all, the whole point of cryptography is to hide
information---and in particular, most cryptographic definitions say
that if an agent receives message $m_1$ encrypted with a key
$k_1$ that she does not know and message $m_2$ encrypted with key
$k_2$ that she also does not know, then that agent should be unable to
distinguish the two messages, in the sense of being able to identify
which is which. 
Thus, goes the argument, a state where an agent has received
$\sencr{m_1}{k_1}$ and a state where that agent has
received $\sencr{m_2}{k_2}$ instead should be indistinguishable if
$k_1$ and $k_2$ are not known. 

To capture a more appropriate definition of state
indistinguishability, one approach is to filter the local states
through a function that replaces all messages encrypted with an
unknown key by a special token $\Box$. More precisely, if $H$ is a set
of messages, we write $[H] = \{[m]^H : m\in H\}$, where $[m]^H$ is
inductively defined as follows:
\begin{align*}
  [m]^H & = m \qquad \text{if $m$ is a plaintext}\\\displaybreak[0]
  [(m_1,m_2)]^H & = ([m_1]^H,[m_2]^H) \\\displaybreak[0]
  [\sencr{m}{k}]^H & = \begin{cases}
      \sencr{[m]^H}{k} & \text{if $k\in\mi{Analyzed}(H)$}\\
      \Box & \text{otherwise}
  \end{cases}
\end{align*}
The revised equivalence relations $\sim^\Box_i$ through which knowledge
is interpreted can now be defined to be $s\sim^\Box_i t$ if and only
if $[\mi{Messages}_i(s)]=[\mi{Messages}_i(t)]$.\footnote{This definition does not
  account for the possibility that an agent, even if she does not know
  the content of an encrypted message, may still recognize that she
  has already seen that encrypted message. (This is an issue when
  encryption is deterministic, so that encrypting $m$ with key $k$
  always yields the same string of bits.) One approach is to refine
  the definition so that every encryption $\sencr{m}{k}$ is replaced by
  a unique token $\Box_{m,k}$.}

The definition $[-]^H$ above, which is typical, uses
$\mi{Analyzed}(-)$ to extract the keys that the agent knows. 
Alternate definitions can be given, from a simpler definition that
looks for keys appearing directly in the local state, to a more
complex recursive definition defined using possible-worlds knowledge.

Possible-worlds knowledge interpreted via an $\sim_i^\Box$ accessibility
relation is
general enough to express message knowledge. 
If we assume a class of propositions $\mi{part}_i(m)$ as
before, true at a state $s$ when $m\in\mi{Parts}(\mi{Messages}_i(s))$,
then formula $K_i(\mi{part}_i(m))$ says that
agent $i$ knows message $m$---intuitively, she knows that $m$ is part
of some message in her local state, and has access to it. 

To see that $K_i(\mi{part}_i(m))$ corresponds to message knowledge as
defined above, we can relate it to the definition of message knowledge
in terms of a local $\mi{knows}_i(m)$ proposition, using the \emph{can
  see} interpretation of message knowledge.
It is not difficult to show that 
if  
$\mi{knows}_i(m)$ is true at state $s$, then 
$K_i(\mi{part}_i(m))$ must also be true at state $s$. 
The converse direction requires a suitable richness condition that
guarantees that there are enough encrypted messages to
compare:
for every message 
$\sencr{m}{k}$ received or intercepted where $k$ is not known to the
agent, there should exist another state in which the agent has
received $\sencr{m'}{k'}$ for a different $m'$ and a different $k'$.\footnote{To see the need for a richness condition, if there
  is a single state in which agent $i$ has received an encrypted
  message, then $K_i(\mi{part}_i(m))$ holds vacuously when $m$ is the
  content of the encrypted message.}
Under such a richness condition, if $K_i(\mi{part}_i(m))$ is true at a
state $s$, then $\mi{knows}_i(m)$ is true at that same state $s$.


Thus, possible-worlds knowledge can be used to express message
knowledge, and can also capture higher-order knowledge, that is,
knowledge about general facts, including other agents' knowledge. 
The informal analyses of \secref{crypto} show that it makes sense to
state that Alice may know that Betty knows the key. While Betty's
knowledge here is message knowledge (knowledge of the key) and
therefore can be modeled with any of the approaches above, Alice's
knowledge is higher-order knowledge, knowledge about knowledge of
another agent.
Logics that allow reasoning about Alice's knowledge of Betty's
knowledge of the key tend to rely on possible-worlds
definitions of knowledge.


\subsection{Reasoning about Cryptographic
  Protocols}\label{chap13:reasoning}

Several approaches have been developed for reasoning about cryptographic
protocols. 
Most are not based on epistemic logic, but extend a classical
propositional or first-order logic---possibly with temporal
operators---with a simple form of message knowledge in the spirit of 
$\mi{knows}_i(m)$.  
This allows them to leverage 
well-understood techniques for system analysis from the formal
verification community and from the programming language community. 
Other approaches are explicitly epistemic in nature.

Techniques for reasoning about cryptographic protocols roughly
split along two axes, each corresponding to a way of using 
logic to reason about protocols in general.
\begin{enumerate}

\item[(1)] Reasoning can be performed either deductively using
  the proof theory of the logic (e.g., through deductions in a theorem
  prover), or semantically, using the models of the logic (e.g., through
  model checking). 

\item[(2)] Reasoning can be performed either directly on the
  description of the protocol---either taken as a sequence of messages
  or a program for each role in the protocol---or indirectly on the
  set of traces generated by protocol executions.

\end{enumerate}
Comparing reasoning methods across these axes is difficult,
as each have their advantages and their disadvantages.

\paragraph{The Inductive Method.}

A good example of a deductive approach for reasoning about security
protocols is the Inductive Method, based on inductive definitions in
higher-order logic (a generalization of first-order predicate logic
allowing quantification over arbitrary relations).
These inductive definitions admit powerful
induction principles which become the main proof technique used to
establish confidentiality and authentication properties. 

The Inductive Method is fairly characteristic of many deductive
approaches to cryptographic protocol analysis: the deductive system is
embedded in a powerful logic such as higher-order logic, and does not
use epistemic concepts beyond a local definition of message knowledge
equivalent to the use of a $\mi{knows}_i(m)$ proposition.

The Inductive Method proper is based on defining a theory---a set of
logical rules---for analyzing a given protocol. The theory
for a protocol describes how to generate the protocol
execution traces, where a trace is a sequence of events
such as \textit{$A$ sends $m$ to $B$}, represented by the predicate
$\mathsf{Say}(A,B,m)$.
Rules state
which events can possibly follow a given sequence of events, thereby
describing traces inductively.
In general, there is a rule in the theory for every message
in the protocol description.
Rules inductively define a set $\mathsf{Prot}$ of traces
representing all the possible traces of the protocol. 

If we consider a theory for
Protocol~\eqref{needham-schroeder-lowe}, 
a rule for message (1) would say: 
\begin{align*}
 & \mi{tr} \in \mathsf{Prot} \\
 & \qquad \rimp 
   \fby{\mi{tr}}{\textsf{Say}(A,B,\sencr{A,n_A}{\mi{pk}_B})}\in \mathsf{Prot}
\end{align*}
where $\fby{\mi{tr}}{e}$ adds event $e$ to trace $\mi{tr}$.
That is, if $\mi{tr}$ represents a valid trace of the protocol, then
that trace can be extended with the first message of a new protocol
execution.
Similarly, a rule for message (2) would say: 
\begin{align*}
& \mi{tr} \in \mathsf{Prot} \\
& \land \mathsf{Say}(A',B,\sencr{A,n_A}{\mi{pk}_B})\in\mi{tr} \\
& \qquad \rimp 
\fby{\mi{tr}}{\mathsf{Say}(B,A,\sencr{B,n_A,n_B}{\mi{pk}_A})}\in\mathsf{Prot}
\end{align*}
That is, if $\mi{tr}$ is a valid trace of the protocol in which an agent
has received the first message of a protocol execution, that agent
can respond appropriately with the second message of the protocol
execution.\footnote{These rules are simplifications. Actual rules
  would contain appropriate quantification and additional side
  conditions to ensure that $A$ and $B$ are different agents, that
  nonces do not clash, and so on.}
Rules are simply implications and conjunctions over a
vocabulary of events. 

The attacker $S$ is also defined by rules; these rules describe how
attacker actions can extend traces with new events. 
For a Dolev-Yao attacker, these rules define a  nondeterministic
process that can intercept any message, decompose it into parts and
decrypt it if the correct key is known, and that can create new
messages from other messages it has observed. 
The theory includes inductive definitions for the $\mi{Analyzed}$ and
$\mi{Synthesized}$ sets given in \secref{knowledge}, as well as rules
of the form
\begin{align*}
 & \mi{tr}\in\mathsf{Prot}\\
 & \land m\in\mi{Synthesized}(\mi{Analyzed}(\mathsf{Spied}(\mi{tr})))\\
 & \land B\in\mathsf{Agents}\\
 & \qquad \rimp 
   \fby{\mi{tr}}{\mathsf{Say}(S,B,m)} \in \mathsf{Prot}
\end{align*}
that states that if $m$ can be synthesized from
the messages the attacker observed on trace $\mi{tr}$ (captured by an
inductively-defined set $\mathsf{Spied}(\mi{tr})$), then the
attacker can add an event $\mathsf{Say}(S,B,m)$ for any agent $B$ to
the trace.

The Inductive Method is geared for proving safety properties: for
every state in every trace, that state is not a bad state.
A protocol is proved correct by induction on the length of the traces: 
choosing the shortest sequence to a bad state, 
assuming all states earlier on the trace are good, then deriving a
contradiction by showing that any state following these good states
must be good itself.

A confidentiality property such as \emph{the attacker never learns
  message $m$} is established by making sure that the attacker is
unable to ever send message $m$, by proving the
following formula:
\[
  (\forall\mi{tr}\in\mathsf{Prot})\,(\forall B\in\mathsf{Agents})\,\mathsf{Say}(S,B,m)\not\in\mi{tr}
\]
This is a \emph{can send} interpretation of message knowledge. 
Indeed, according to the rules for the attacker, if the attacker
knows message $m$ at any point during a trace, then there exists a
extension of that trace where the attacker sends message $m$. 
Thus, showing that the attacker never learns message $m$ amounts to
showing that there is no trace in which an event
$\mathsf{Say}(S,B,m)$ appears, for any agent $B$. 

Abstracting away from the details of the approach, the Inductive
Method essentially relies on rules to describe the evolution of a
protocol execution, and verifying a confidentiality property is reduced
to verifying that a certain bad state is not reachable. Other
approaches to cryptographic protocol analysis share this methodology,
many of them using a logic programming language rather than
higher-order logic to express protocol evolution rules; see
\secref{bibnotes}.

\paragraph{BAN Logic.}
The Inductive Method relies on encoding rules for generating protocol
execution traces in an expressive general logic suitable for automating
inductive proofs. 
In contrast, the next approach, BAN Logic, is a logic tailored for
reasoning about cryptographic protocols described as a sequence of
message exchanges. 
It has the additional feature of including a higher-order
\emph{belief} operator as a primitive.

BAN Logic is a logic in the tradition of Hoare Logic, in that
it advocates an axiomatic approach for reasoning about
cryptographic protocols. 
BAN Logic tracks the evolution
of beliefs during the execution of cryptographic protocol, and is
described by a set of inference rules for deriving new beliefs from
old.
BAN Logic includes primitive formulas stating that $k$ is a 
shared key known only to $A$ and $B$ ($A\key{k}B$), that $m$ is a
secret between  
$A$ and $B$ ($A\secret{m}B$), that agent $A$ believes 
formula $F$ ($A\believes F$), that agent $A$ controls the truth 
of formula $F$ ($A\controls  F$), that agent $A$ sent a message
meaning $F$ ($A\said F$), that agent $A$ received and understood
a message meaning $F$ ($A\sees F$), and that a message meaning $F$ 
was created during the current protocol execution ($\fresh( F)$).  
The precise semantics of these formulas is given indirectly through
inference rules, some of which are presented below.

BAN Logic assumes that agents can recognize
when an encrypted message is one they have created
themselves; encryption is in consequence written $\sencr{F}{k}^i$,
where $i$ denotes the agent who encrypted a message meaning $F$ with
key $k$. (This also highlights another characteristic of BAN Logic:
messages are formulas.)

Here are some of the inference rules of BAN Logic:
\begin{align*}
& \text{(R1)} & 
 & \Rule{A\believes B\key{k}A\quad A\sees\{ F\}^i_k \quad i\ne A}
       {A\believes B\said  F} \\\displaybreak[0]
& \text{(R2)} &
 & \Rule{A\believes B \said (F,F')}{A \believes B\said F}\\\displaybreak[0]
& \text{(R3)} & 
 & \Rule{A\believes\fresh( F)\quad A\believes(B\said  F)}
       {A\believes B\believes  F}\\\displaybreak[0]
& \text{(R4)} &
 & \Rule{A\believes B\controls  F \quad A\believes B\believes  F}
         {A\believes  F} \\\displaybreak[0]
& \text{(R5)} &
 & \Rule{A\sees ( F, F')}{A\sees  F}\\\displaybreak[0]
& \text{(R6)} &
 & \Rule{A\believes B\key{k}A\quad A\sees\{ F\}^i_k \quad i \ne A}{A\sees  F}\\\displaybreak[0]
& \text{(R7)} &
 & \Rule{A\believes \fresh( F)}{A\believes \fresh (( F', F))}\\\displaybreak[0]
& \text{(R8)} & 
 & \Rule{A \believes B \believes ( F, F')}
        {A \believes B \believes  F}
\end{align*}
Rule (R1), for instance, says that if agent $A$ believes that $k$ is
shared only between $B$ and herself, and she receives a message
encrypted with 
key $k$ that she did not encrypt herself, then she believes that $B$
sent the original message.
Rule (R3) is an honesty rule: it says that agents send messages
meaning $F$ only when they believe $F$. 
There are commutative variants of rules (R2), (R5), (R7), and
(R8), as well as variants for more general tuples; there are also 
variants of (R8) for any level of nested belief.

BAN Logic does not attempt to model protocol execution traces.
Reasoning is done directly on the sequence of messages in the
description of the protocol. 
Because sequences of messages do not 
carry enough information to permit this kind of reasoning, 
a transformation known as \emph{idealization} must be applied to the protocol. 
Roughly speaking, idealization consists of
replacing the messages in the protocol by formulas of BAN Logic that
capture the intent of each message. For
instance, if agent $A$ sends key $k$ to agent $B$ with the
intention of sharing a key that is known only to $A$, then a suitable
idealization would have $A$ send the formula
$A\key{k}B$ to $B$. 
Idealization is an annotation mechanism, and as such
is somewhat subjective. 

To illustrate reasoning in BAN Logic, consider the following simple
protocol in which Alice sends a secret value $m_0$ to Betty encrypted
with their shared key $k_{\mi{AB}}$, along with a nonce exchange to
convince $B$ that the message is not a replay of a message in a
previous execution of the protocol
(see \secref{crypto}):
\begin{protocol}{prot-shared}
\pmessage{1}{A}{B}{A}\\
\pmessage{2}{B}{A}{n_B}\\
\pmessage{3}{A}{B}{A,\sencr{m_0,n_B}{k_{\mi{AB}}}}
\end{protocol}%
A possible 
idealization of protocol~\eqref{prot-shared} would be:
\begin{protocol}{prot-shared-ideal}
\pmessage{3'}{A}{B}{\sencr{A\secret{m_0}B,n_B}{k_\mi{AB}}}
\end{protocol}%
The first two messages in protocol~\eqref{prot-shared} carry
information that BAN Logic does not use, so they are not present in the
idealized protocol. The third message is idealized to $A$ sending
formula $A\secret{m_0}B$ to $B$ along with the nonce $n_B$,
indicating that $A$ considers $m_0$ to be a secret at that point.

Reasoning about an idealized protocol consists in laying out the
initial beliefs of the agents, and deriving new beliefs from those and
from the messages exchanged between the agents, using the inference
rules of the logic. For protocol~\eqref{prot-shared-ideal}, 
initial beliefs include that both parties believe that key
$k_\mi{AB}$ has not been compromised, that
nonce $n_B$ has not already been used, and that message
$m_0$ that $A$ wants to send to $B$ is initially secret.
These initial beliefs are captured by the following formulas:
\begin{align*}
 &  A\believes A\key{k_\mi{AB}}B \\
 &  B\believes A\key{k_\mi{AB}}B \\
 &  B\believes\fresh(n_B)\\
 &  A\believes A\secret{m_0}B
\end{align*}
We can derive new formulas from these initial beliefs in
combination with the messages exchanged by the agents. The idea is to
update this set of formulas after each protocol step: after an
idealized step $A\rightarrow B: F$, which says that $B$ receives a
message meaning $F$,  we can add formula $B\sees F$ to the set
of formulas, and we use the inference rules to derive additional formulas
to add to the set.

For example, in idealized protocol~\eqref{prot-shared-ideal}, after
message (3'), we add formula
\begin{equation}\label{e:Bseesencryption}
 B \sees \sencr{A\secret{m_0}B,n_B}{k_{\mi{AB}}}
\end{equation}
to the set of initial beliefs.
Along with the initial belief $B\believes
A\key{k_\mi{AB}}B$, formula~\eqref{e:Bseesencryption} allows us to apply inference rule (R1) to derive:
\begin{equation}\label{e:BbelievesAsaid}
B\believes A\said (A\secret{m_0}B,n_B).
\end{equation}
From the initial belief $B\believes\fresh(n_B)$, inference rule (R7)
lets us derive that any message combined with $n_B$ must be fresh,
and thus we can derive:
\begin{equation}\label{e:Bbelievesfresh}
  B\believes\fresh(A\secret{m_0}B,n_B).
\end{equation}
Formula~\eqref{e:Bbelievesfresh} together with
\eqref{e:BbelievesAsaid} give us, via inference rule (R3): 
\[
B\believes A\believes (A\secret{m_0}B,n_B).
\]
Via inference rule (R8), this yields:
\begin{equation}\label{e:BbelievesAbelieves}
B\believes A\believes A\secret{m_0}B.
\end{equation}
Thus, after the messages of the idealized protocol have been
exchanged, $B$ believes that $A$ believes that $m_0$ is a secret
between $A$ and $B$.  This is about as much as we can expect. 

We can say more if we are willing to assume that $B$ believes that the
secrecy of $m_0$ is in fact controlled by $A$. If so, we can add the
following formula to the set of initial beliefs:
\begin{equation}\label{e:BbelievesAcontrols}
  B\believes A\controls A\secret{m_0}B
\end{equation}
and formulas~\eqref{e:BbelievesAcontrols} and \eqref{e:BbelievesAbelieves}
combine via inference rule (R4) to yield the stronger conclusion:
\[ B\believes  A\secret{m_0}B.\]   
In other words, if $B$ believes that $A$ controls the secrecy of $m_0$
and also that $A$ believes $m_0$ to be secret, then after the protocol
executes $B$ also believes that $m_0$ is a secret shared only with
$A$.  

Attackers are not explicit in BAN Logic.
In a sense, an active Dolev-Yao attacker is implicitly encoded within
the inference rules of the logic, but the focus of BAN Logic is
reasoning about the belief of agents in the presence of an active 
attacker, as opposed to reasoning about the knowledge of an attacker.
A successful attack in BAN Logic shows up as a failure to establish a
desired belief for one of the agents following a protocol execution.


\paragraph{Temporal and Epistemic Temporal Logics.}

Another class of approaches for reasoning about cryptographic
protocols rely on a form of temporal logic to express desired
properties of the protocol and show that they are true of a model
representing the protocol---generally
through a suitable representation of its traces. 
This is done through
model-checking techniques to determine algorithmically whether a
formula is true in the models representing the protocol. 
These model-checking techniques vary in terms of how the models are
described: these can be either directly expressed by finite state machines,
or through domain-specific languages.

The simplest approaches to cryptographic protocol analysis via
temporal logics merely extend existing temporal-logic verification
techniques.
At least two challenges arise in these cases: modeling attackers, and
expressing message knowledge. 
For attackers, while eavesdropping attackers do not affect the
execution of protocols and therefore are comparatively easy to handle
in standard temporal-logic verification frameworks, active attackers
require work. In some cases, it is possible to simply encode an active
attacker within the model using the tools of the framework. Message
knowledge is usually dealt with by introducing a variant of a
$\mathit{knows}_i(m)$ proposition.  

In general, the logics themselves are completely straightforward: they
are standard propositional or first-order temporal logics extended
with a message knowledge predicate. All the action is in the
interpretation of the message knowledge predicate, and in the
construction of the models to account for the actions of active
attackers. 
There is not much to say about those approaches as far as pertains to
epistemic concepts, but they are popular in practice. 

More interesting from an epistemic perspective are those frameworks
relying on a temporal epistemic logic, that is, a logic with both
temporal and epistemic operators. 
The MCK model-checker is an example of a verification framework that
uses a linear-time temporal logic with
epistemic operators to verify protocols that do not use cryptography,
such as the Seven Hands or the Dining Cryptographers protocols
of \secref{protocols}. 
Protocols are described via finite state machines, and formulas express
properties of paths through that finite state machine, each such path
corresponding to a possible execution of the protocol. 

As an example, consider the Dining Cryptographers protocol, which
translates well to a finite state machine. States can be described
using three 
agent-indexed Boolean variables $\mi{paid}[i]$, $\mi{chan}[i]$, and
$\mi{df}[i]$, where variable $\mi{paid}[i]$ records whether agent $i$
paid;, variable $\mi{chan}[i]$ is a communication channel used
by agent $i$ 
to send the result of its coin toss to her right neighbor, and
variable $\mi{df}[i]$ records the announcement of $\mi{Df}_i$ by agent
$i$ at the end of the protocol. The initial states are all the states
satisfying: 
\begin{align*}
& (\neg paid[1] \land \neg paid[2] \land \neg paid[3]) \lor (paid[1]
  \land \neg paid[2] \land \neg paid[3])\\ 
 & \quad \lor (\neg paid[1] \land paid[2] \land \neg paid[3]) \lor (\neg
  paid[1] \land \neg paid[2] \land paid[3]) 
\end{align*}
Every agent executes the following program, where a single step of the
program for each agent is executed in a transition of the state 
machine: 
\begin{verbatim}
    protocol diningcrypto (paid : observable Bool, 
                           chan_left, chan_right : Bool, 
                           df : observable Bool[])

      coin_left, coin_right : observable Bool

      begin
        if    True -> coin_right := True 
           [] True -> coin_right := False 
        fi;
        chan_right.send(coin_right);
        coin_left := chan_left.recv();
        df[self] := coin_left xor coin_right xor paid;
      end
\end{verbatim}
Program \texttt{diningcrypto}\footnote{The \texttt{observable} annotation is
  used to derive the indistinguishability relation: two states are
  indistinguishable to agent $i$ if the observable variables of the
  program executed by agent $i$ have the same value in both
  states. The \text{\texttt{if} 
  ... \texttt{[]} ... \texttt{fi}} construct nondeterministically
  executes one of its branches with an associated condition that
  evaluates to true. Variable \texttt{self} is assigned the
  name of the agent executing the program.} is instantiated 
for every agent with suitable variables for the parameters:
\begin{align*}
& \text{agent 1 executes \texttt{diningcrypto ($\mi{paid}[1]$,$\mi{chan}[1]$,$\mi{chan}[2]$,$\mi{df}$)}}\\
& \text{agent 2 executes \texttt{diningcrypto ($\mi{paid}[2]$,$\mi{chan}[2]$,$\mi{chan}[3]$,$\mi{df}$)}}\\
& \text{agent 3 executes \texttt{diningcrypto ($\mi{paid}[3]$,$\mi{chan}[3]$,$\mi{chan}[1]$,$\mi{df}$)}}
\end{align*}
At the first state transition, every agent nondeterministically chooses
a value for their coin toss into local variable \texttt{coin\_right}; at the second state transition, the
result of the coin toss is sent on the channel given as the
\texttt{chan\_right} parameter; at the third state transition, the
local variable \texttt{coin\_left} for each agent is updated to
reflect the result of the coin toss received from the agent's left
neighbor; at the fourth state transition, variable $\mi{df}$ is
updated for every agent.

Given such a state machine, a formula expressing the
anonymity of the 
payer from agent 1's perspective can be written as:
\[
   \mathbf{X}^4\begin{aligned}[t]& ( \neg \mi{paid}[1]\\
  & \qquad \rimp (K_1 (\neg \mi{paid}[1] \land \neg \mi{paid}[2] \land \neg \mi{paid}[3])) \\
  & \qquad \qquad \lor (K_1 (\mi{paid}[2] \lor \mi{paid}[3])
\land \neg K_1 \mi{paid}[2] \land \neg K_1 \mi{paid}[3]))
   \end{aligned}
\]
where $\mathbf{X}^4$ is a temporal operator meaning \emph{after four
  rounds}. 
This formula, which is true or false of an initial state,
says that after the protocol terminates, if
cryptographer 1 did not pay, then she either knows that no
cryptographer paid, or she knows that one of the other two
cryptographers paid but does not know which. 
(Formulas expressing anonymity from agent 2 and agent 3's perspectives
are similar.)

MCK has no built-in support for active attackers, so it cannot easily
deal with cryptographic protocols even if we were to add a message
knowledge primitive to the language that can deal with encrypted
messages.
Of course, it is possible to encode some attackers within the language
that MCK provides for describing models, but the effect on the
efficiency of model checking is unclear.

The theoretical underpinnings of model checking for temporal epistemic
logic are fairly well understood, even though the problem has not been
studied nearly as much as model checking for temporal logics. 
Message knowledge does not particularly complicate matters, once
the choice of how to interpret message knowledge is made. 
Accounting for active attackers is more of an issue, since active
attackers introduce additional actions into the model, increasing its
size. 

The main difficulty with model checking epistemic temporal logic is
its inherent complexity. 
While model checking a standard epistemic logic such as S5 takes time
polynomial in the size of the model, adding temporal operators and
interpreting the logic over the 
possibly infinite paths in a finite state machine increases that
complexity. 
For example, in the presence of perfect recall (when agents remember
their full history) and synchrony (when agents have access to a
global clock), the model-checking problem has non-elementary
complexity if the logic includes an \emph{until} temporal operator, and is
PSPACE-complete otherwise. 
The problem tends to become PSPACE-complete when perfect recall
is dropped. 
Progress has been made to control the complexity of
model-checking epistemic temporal logics by a 
careful analysis of the complexity of specific classes of formulas
that, while restricted, are still sufficiently expressive to capture
interesting security properties, but much
work remains to be done to make the resulting techniques efficient.


\section{Information Flow in Multi-Level Systems}
\label{chap13:iflow}

Confidentiality in cryptographic protocols is mainly viewed through the
lens of access control: some privilege (a key) is required in order
to access the confidential data (the content of an encrypted
message). 
An agent who has the key can access the content, an agent without the
key cannot. 
Those access restrictions can control the release of information, but
once that information is released there is nothing stopping it from
being propagated by agents or by the system through error or malice,
or because the released information is needed for the purpose of
computations.   
For systems in which confidentiality is paramount, it is not
sufficient to simply ensure that access to confidential data is
controlled, there also needs to be a guarantee that even when the
confidential data is released it does not land in unauthorized hands.
These sorts of confidentiality guarantees require understanding the
flow of information in a system.

Confidentiality in the presence of released information is usually
studied in the context of systems in which all data are classified
with a security 
level, and where agents have security clearances allowing
them to access data at their security level or lower. 
For simplicity, only scenarios with security
levels \emph{high} and \emph{low} (think \emph{classified} and
\emph{unclassified} in military settings) will be considered.
Intuitively, a high-security agent should be allowed to read both high-
and low-security data, and a low-security agent should be allowed to read
only low-security data. 
This is an example of security policy, which describes the forms of
information flows that are allowed and those that are
disallowed. Information flows that are disallowed capture the desired
confidentiality guarantee.

As an example, imagine a commercial system such as a bank mainframe,
where agents perform transactions via credit cards or online
accounts. In such a system, credit card numbers and bank account
numbers might be considered high-security data, and low-security
agents should be prevented from accessing them.
However, what about the last four digits of a credit card number? 
Even that information is often considered sensitive. What about a
single digit? What about the digits frequency in any given credit card
number?
Because it is in general difficult to characterize exactly what kind
of information about high-security data should not be leaked to
low-security agents, it is often easier to prevent any kind of partial
information disclosure.

The problem of preventing information disclosure is made more interesting,
and more complicated, by the fact that information may not only flow
directly from one point to another (e.g, by an agent sending a message
to another, or by information being posted, or by updating an
observable memory location) but may also flow indirectly from one
point to another. 
Suppose that the commercial system described above sends an email to
a central location whenever a transfer of more than one million
dollars into a given account $A$ occurs. 
Anyone observing email traffic can see those emails being sent and
learn that account $A$ now contains at least a million dollars. 
This is an extreme example, but it illustrates indirect information
flow: information is gained not by directly observing an event, but by
correlating an observation with the event. 

Epistemic concepts arise naturally in this setting---a security policy
saying that there is no flow of information from high-security data to
low-security agents can be expressed as \emph{low-security agents do
  not learn anything about high-security data}. Moreover, the
definitions used in the literature essentially rely on a
possible-worlds definition of knowledge within a specific class of
models. 

Two distinct models of information flow will
be described. 
Both of these models are observational models: they define the kind of
observations that agents can make about the system and about the
activity of other agents. 
These observations form the basis of agents' knowledge. 

The first model considered takes a fairly abstract view of a system,
as sequences of events such as inputs from agents, outputs to agents,
internal computation, and so on. These events are the observations
that agents can make. In such a setting, 
security policies regulate information about the occurrence of events. 
The second model considered is more concrete, and stems from practical
work on defining verification techniques for information flow at the
level of the source code implementing a system. In that model,
observations take the form of content of memory locations that
programs can manipulate.


\subsection{Information Flow in Event Systems}
\label{chap13:events}

The first model of information flow uses sets of traces corresponding
to the possible executions of the system.
Every trace is a sequence of events; some of those
events are high-security events (and only observable by high-security
agents), and some of those events are low-security events (and
observable by both high-security and low-security agents).
The intuition is that a low-security agent, observing only the low-security
events in a trace, should not be able to infer any information about
the high-security events in a trace. 

How can a low-security agent infer information?
If we assume that the full set of traces of the system is known to all
agents, then a low-security agent, upon observing a particular
sequence of low-security events, can narrow down a set of possible
traces that could be the actual trace by considering all the traces
that are compatible with her view of the low-security events. 
By looking at those possible traces, she may infer information about
high-security events. For instance, maybe a particular high-security
event $e$ appears in every such possible trace, and thus she learns
that high-security event $e$ has occurred.
In the most extreme case, there may be a single trace compatible with
her view of the low-security events, and therefore that low-security
agent learns exactly which high-security events have occurred. 

The model can be formalized using event systems. 
An event system is a tuple $S=(E,I,O,\mi{Tr})$ where $E$ is a
set of events, $I\subseteq E$ a set of input events,
$O\subseteq E$ a set of output events, and $\mi{Tr}\subseteq E^*$ a
set of finite traces representing the possible
executions of the system.
Given a trace $\tau\in E^*$ and a subset $E'\subseteq E$ of
events, we write $\tau\restr{E'}$ for the subtrace of $\tau$
consisting of events from $E'$ only.

We assign a security level to events in $E$ by partitioning them into
low-security events $L$ and high-security events $H$: 
events in $I\cap L$ are low-security input events, events in
$O\cap L$ are low-security output events, and so on. 

A naive attempt at defining information flow in this setting might be to
say that there is information flowing from high-security events to
low-security agents if a low-security agent's view of $\tau\restr{L}$
implies that at least one high-security event subsequence 
is not possible. In other words, seeing a particular sequence of
low-security events rules out one possible high-security event
subsequence. 
Formally, if we write $\mi{Tr}\restr{H}$ for
$\{\tau\restr{H}:\tau\in\mi{Tr}\}$, information flows from
high-security events to low-security agents if there is a trace
$\tau\in\mi{Tr}$ such that $\{\tau'\restr{H} :
\tau'\restr{L}=\tau\restr{L}\}\ne\mi{Tr}\restr{H}$. 

%

Such a definition turns out to be too strong---it is
equivalent to separability described below---because it
pinpoints information flows where there are none: since
low-security events may influence high-security events, a
particular subsequence of high-security events may be ruled out due to the
influence of low-security events, and in that case there should be no
information flow since the low-security agent could have already predicted
that the high-security subsequence would have been ruled out. 
Intuitively, there is information flow when one high-security event
subsequence that should be possible as far as the low-security agent
expects is not in fact possible.
This argument gives an inkling as to why the definition of information
flow is not entirely trivial.

Security policies in event systems are often defined as closure
properties of the set of traces. Security policies 
that historically were deemed interesting for the purpose of
formalizing existing multi-level systems include the following: 

\begin{itemize}

\item \textbf{Separability:} 
no nontrivial interaction between high-security
  events and low-security agents is possible because for any such interaction
  there is a trace with the same high-security events but different
  low-security events, and a trace with the same low-security events
  but different high-security events. Formally,
  for every pair of traces $\tau_1,\tau_2\in\mi{Tr}$, 
  there is a trace $\tau\in\mi{Tr}$ such that
  $\tau\restr{L}=\tau_1\restr{L}$ and
  $\tau\restr{H}=\tau_2\restr{H}$. 

\item \textbf{Noninference:} 
a low-security agent cannot
learn about the occurrence of high-security events because any 
trace, as far as the low-security agent can tell, could be a trace
where there are no high-security events at all. Formally,
for all traces $\tau\in\mi{Tr}$, there is a
trace $\tau'\in\mi{Tr}$ such that $\tau\restr{L}=\tau'\restr{L}$ and
$\tau'\restr{H}=\langle\,\rangle$. 

\item \textbf{Generalized Noninference:} 
A more lenient form of noninference, 
where a low-security agent cannot
learn about the occurrence of high-security input events because any
trace could be a trace
where there are no high-security input events at all. Formally,
for all traces $\tau\in\mi{Tr}$, there is
  a trace $\tau'\in\mi{Tr}$ such that $\tau\restr{L}=\tau'\restr{L}$ and
  $\tau'\restr{(H\cap I)}=\langle\,\rangle$. 

\item \textbf{Generalized Noninterference:} a low-security agent
  cannot learn about high-security input events, and 
  high-security input events cannot influence low-security
  events. Formally, for all traces $\tau\in\mi{Tr}$ and all traces
  $\tau'\in\mi{interleave}((H\cap I)^*,\{\tau\restr{L}\})$, there is a
  $\tau''\in\mi{Tr}$ such that $\tau''\restr{L}=\tau\restr{L}$ and
  $\tau''\restr{(L\cup(H\cap I))}=\tau'$. (Function
  $\mi{interleave}(T,U)$ returns every possible interleaving
  of every trace from $T$ with every trace from $U$.)

\end{itemize}

A closure property says that if some traces are in the model,
  then other variations on these traces must also be in the model.
This is clearly an epistemic property. 
Under a possible-worlds definition of knowledge, an agent knows a
formula if that formula is true at all
traces that the agent considers possible given her view of the
system.
In general, the fewer possible traces there are, the more
facts can be known, since it it easier for a fact to be true at all
possible traces if there are few of them. 
The closure properties ensure that there are enough possible traces
from the perspective of a low-security agent to prevent a specific of
class facts from being known.\footnote{As in \secref{cryptoprotocols},
 this does not take probabilistic information into account.} 

Closure conditions on sets of traces are therefore just a way to
enforce lack of knowledge, given a possible-worlds definition of
knowledge. 
We can make this precise by viewing event systems as
Kripke frames. 

The accessibility relation of each agent depends on the agent's security
level. In the case of interest, a low-security agent is assigned an
accessibility relation $\sim_L$ defined as $\tau_1\sim_L \tau_2$ if
and only if $\tau_1\restr{L}=\tau_2\restr{L}$.

We identify a proposition with a set of traces, intuitively, those
traces in which the proposition is true.
As usual, conjunction is intersection of propositions, disjunction is
union of propositions, negation is complementation of propositions
with respect to the full set of traces in the event system, and
implication is subset inclusion.
To define the proposition \emph{the low-security agent knows $P$}, we
first define the low-security agent's knowledge set of a trace $\tau$ as the set of all
traces $\sim_L$-equivalent to $\tau$, $\cK_L(\tau) = \{\tau' :
\tau\sim_L\tau'\}$.
The proposition \emph{the low-security agent knows $P$} can be defined in
the usual way, as: 
\[ \sK_L(P) = \{\tau : \cK_L(\tau)\subseteq P\} \]
It is easy to see that $\sK_L$ satisfies the usual S5 axioms, suitably
modified to account for propositions being sets:
\begin{align*}
  \textrm{(D)}  \qquad &  \sK_L(P) \cap \sK_L(Q) = \sK_L(P\cap Q)\\
  \textrm{(K)}  \qquad & \sK_L(P) \subseteq P\\
  \textrm{(PI)} \qquad & \sK_L(P) \subseteq \sK_L(\sK_L(P))\\
  \textrm{(NI)} \qquad & \neg\sK_L(P) \subseteq \sK_L(\neg\sK_L(P))
\end{align*}
These properties are the set-theoretic analogues of
\emph{Distribution}, \emph{Knowledge}, \emph{Positive Introspection},
and \emph{Negative Introspection}, respectively.

As an example, consider an event system $(E,I,O,\mi{Tr})$ that satisfies
generalized noninterference, and the proposition 
\emph{high-security input event $e$ has occurred}. 
This proposition is represented by the set $P_e$ of all traces in
$\mi{Tr}$ in which $e$ occurs.
The proposition \emph{the low-security agent knows that $e$ has occurred}
corresponds to the set of traces $\sK_L(P_e)$. 
It is easy to check that because the system satisfies generalized
noninterference, the set $\sK_L(P_e)$ is empty, meaning that there
is no trace on which the low-security agent ever knows $P_e$,
that is, that $e$ has occurred.  
By way of contradiction, suppose that $\tau\in\sK_L(P_e)$. 
By definition, $\tau\in\sK_L(P_e)$ if and only if
$\cK_L(\tau)\subseteq P_e$. 
But the closure condition for generalized noninterference implies that
there must exist a trace $\tau'\sim_L\tau$, that is, a trace in
$\cK_L(\tau)$, such that $e$ does not occur in $\tau'$. Thus,
there is a trace in $\cK_L(\tau)$ which is not in $P_e$, and
$\cK_L(\tau)\not\subseteq P_e$. Thus, $\tau\not\in\sK_L(P_e)$,
a contradiction. 

This is a somewhat roundabout way to see that there is an implicit
epistemic logic lurking which
explains the notions of information flow security policies in event
systems. 
It is certainly possible to make such a logic explicit by introducing
a syntax and adding an interpretation to event systems, and
study information flow in event systems from such a perspective. 

The key point here is that event-system models of information flow
and the expression of security policies in those models intrinsically
use epistemic concepts, and all reasoning 
is essentially classical epistemic reasoning performed directly on the
models.


\subsection{Language-Based Noninterference}
\label{chap13:languages}

A more concrete model for information flow is obtained by moving away
from trace-based models of systems and relying instead on the program
code implementing those systems. 

Defining information flow at the level of programs has several
advantages: the system is described in detail, information can be
defined in terms of the data explicitly manipulated by the program,
and enforcement can be automated; the latter turns out to be
especially important given the complexity of modern computing systems
which makes manual analysis often infeasible. 

The observational model used by most language-based information-flow
security research is not event-based, although it is still broadly
concerned with input and output.
The focus here is on information flow in imperative programs, which
operates by changing the state of the environment as a program executes.
The state of the environment is represented by a store holding values
associated with variables.
Variables can be read and written by programs. 
Every variable is tagged with a security level, describing the security
level of the data it contains.
A low-security agent can observe all low-security variables, but not
the high-security ones. 
Inputs to programs are modeled as initial values of variables,
while outputs are modeled as final value of variables: 
low-security inputs are initial value of low-security variables, and
so on.
The basic security policy generally considered is a form of
noninterference: that low-security outputs do not reveal anything
about high-security inputs, and that high-security inputs do
not influence the value of low-security outputs.

Consider the following short programs:
\begin{align*}
 \text{(P1)} \qquad & h := l + 1 \\
 \text{(P2)} \qquad & l := h +1  \\
 \text{(P3)} \qquad & \ms{if}~l=0~\ms{then}~h:=h+1~\ms{else}~l:=l+1 \\
 \text{(P4)} \qquad & \ms{if}~h=0~\ms{then}~h:=h+1~\ms{else}~l:=l+1
\end{align*}
In all of these programs, variable $h$ is a
high-security variable, and variable $l$ is a low-security
variable. 

A program executes in a store assigning initial values to variables,
and execution steps modify the store until the program terminates in a
final store. 
Several simplifying assumption are made: programs are
deterministic, and programs always terminate. 
This is purely to keep the discussion and the technical machinery
light. 
These restrictions can be lifted easily. 
Moreover, the programming language under consideration will not be
described in detail; the sample programs should be intuitive enough.

How do we formalize noninterference in this setting? 
A store $\sigma$ is a mapping from variables $x$ to values
$\sigma(x)$. 
We assume every variable $x$ is tagged with a security level
$\mi{sec}(x)\in\{L,H\}$. 
Let $\Sigma$ be the set of all possible stores.
We model execution of a program $C$ using a function
$\qq{C}:\Sigma\longrightarrow\Sigma$ from initial stores to 
final stores.
Thus, executing program $C$ in store $\sigma$ yields a final store
$\qq{C}(\sigma)$. 
For example, executing program (P1) in store
$\langle l\mapsto 5,h\mapsto 10\rangle$ yields store $\langle l\mapsto 5,
  h\mapsto 6\rangle$, and executing program (P3) in store
$\langle l\mapsto 5,h\mapsto 10\rangle$ yields store $\langle l\mapsto 6,
  h\mapsto 10\rangle$. 

Two stores $\sigma_1$ and $\sigma_2$ are $L$-equivalent,
written $\sigma_1\approx_L \sigma_2$, if they assign the same values
to the same low-security variables: $\sigma_1\approx_L\sigma_2$ if and
only if for all variables $x$ with $\mi{sec}(x)=L$,
$\sigma_1(x)=\sigma_2(x)$.  
A program $C$
satisfies noninterference if executing $C$ in two $L$-equivalent
states (that is, in two states that a low-security agent cannot
distinguish) yields two $L$-equivalent states: for all $\sigma_1$ and
$\sigma_2$, if $\sigma_1\approx_L\sigma_2$, then
$\qq{C}(\sigma_1)\approx_L\qq{C}(\sigma_2)$.\footnote{Another way of
  understanding this definition is that it requires the
  relation on stores induced by program execution to be a refinement
  of $L$-equivalence $\approx_L$. If we
  define $\qq{C}_{\approx_L}$ as the relation
  $\{(\qq{C}(\sigma_1),\qq{C}(\sigma_2)) : \sigma_1\approx_L\sigma_2\}$,
then the noninterference condition can be rephrased as
$\qq{C}_{\approx_L}\subseteq{}\approx_L$.}

How do programs (P1--4) fare under this definition of
noninterference? Program (P1) clearly satisfies
noninterference, since the final value of low-security variable $l$
does not depend on the 
value of any high-level variable, while program (P2) clearly
does not. The other two programs are more interesting. The final value
of low-security variable $l$ in program (P3) only depends on
the initial value of $l$, and thus we expect (P3) to satisfy
noninterference, and it does. Program (P4), 
however, does not, as we can see by executing the program in stores
$\langle l\mapsto 0, h\mapsto 0\rangle$ and $\langle l\mapsto 0, h\mapsto 1\rangle$,
both $\approx_L$-equivalent, but which yield stores
$\langle l\mapsto 0, h\mapsto 1\rangle$ and $\langle l\mapsto 1,h\mapsto 1\rangle$,
respectively, two stores that cannot be $L$-equivalent since they differ
in the value they assign to variable $l$. And indeed, observing the final value of $l$ reveals information about the initial value of $h$. 

Noninterference is usually established by a static analysis of the
program code, which approximates the flow of information through a
program before execution.
While the details of the static analyses are interesting
in their own right, they have little to do with epistemic logic beyond
providing an approach to verifying a specific kind of epistemic
property in a specific context.

Recent work on language-based information-flow security has
highlighted the practical importance of declassification, that is, the 
controlled release of high-security data to low-security agents.
The problem of password-based authentication illustrates the need for
such release: when a low-security agent tries to authenticate herself
as a high-security agent, she may be presented with a login screen
asking for the password of the high-security agent. That password
should of course be considered high-security information.
However, the login screen leaks information, since entering an
incorrect password will reveal that the attempted password is not the
right password, thereby leaking a small amount of information about
the correct password. The leak is small, but it exists, and because of
it the login screen does not satisfy the above definition of
noninterference. Defining a suitable notion of security policy that
allows such small release of information while still preventing more
important information flow is a complex problem.

While the concepts underlying information-flow security are clearly
epistemic in nature---taking stores as possible worlds and
$L$-equivalence as an accessibility relation for low-security
agents---there is no real demand for an explicit epistemic logic in
which to describe policies.
One reason is that it is in general
difficult to precisely nail down, in a given system, what
high-security information should be kept from low-security agents. 
It is simply easier to ask that no information be leaked to
low-security agents. This \emph{no information} condition is easier to
state semantically than through an explicit logical language---not
learning any information in the sense of noninterference can be stated
straightforwardly as a relationship between equivalence relations,
while if we were to use an epistemic logic, we would have to say
something along the lines of \emph{for all formulas 
  $\phi$ that do not depend only on the state of the low-security
  agent, $\neg K_L\phi$} where $K_L$ expresses the knowledge of that
low-security agent. 
The latter is patently clunkier to work with. It may be the case that
an explicit epistemic logic would be more useful in the context of
declassification, where not all information needs to be
kept from low-security agents.


\section{Beyond Confidentiality}
\label{chap13:beyond}

The focus of this chapter has been on confidentiality, because
it is by far the most studied security property. 
It is not only important, it also underpins several other security
properties. Other related properties are also relevant.

\paragraph{Anonymity.}
A specific form of confidentiality is anonymity,
where the information to be kept secret is the association between
actions and agents who perform them. 
Anonymity has been studied using epistemic logic, and several related
definitions have been proposed and debated. 

To discuss anonymity, we need to be able to talk about actions and 
agents who perform them. Let $\delta(i, a)$ be a proposition
interpreted as \emph{agent $i$ performed action $a$}. 

The simplest definition of anonymity is lack of knowledge: action $a$
performed by agent $i$ is minimally anonymous with respect to agent
$j$ if agent $j$ does not know that agent $i$ performed $a$. This can be
captured 
  by the formula \[\neg K_j\delta(i,a).\]

Minimal anonymity is, well, minimal. It does not rule out that agent
$j$ may narrow down the list of possible agents that performed $a$ to
agent $i$ and one other agent. Stronger forms of anonymity can be
defined: action $a$ performed by agent $i$ is totally anonymous with
respect to agent $j$ if, as far as agent $j$ is concerned, action $a$ could have
been performed by any agent in the system (except for agent $j$). This can be
captured by the formula
\[\delta(i,a)\rimp\bigwedge_{i'\ne j}P_j\delta(i',a)\]
where $P_i\phi$ is the usual dual to knowledge, $\neg K_i(\neg\phi)$,
read as \emph{agent $i$ considers $\phi$ possible}. 

Total anonymity is at the other extreme on the spectrum from minimal
anonymity; it is a very strong requirement. 
Intermediate definitions can be obtained by requiring that actions be
anonymous only up to a given set of agents---sometimes called an
anonymity set: action $a$ performed by agent $i$ is anonymous up to
$I$ with respect to agent $j$ if, as far as agent $j$ is concerned, action $a$
could have been performed by any agent in $I$. This can be
captured by the formula: 
\[\delta(i,a)\rimp\bigwedge_{i'\in I}P_j\delta(i',a).\]

As an example of this last definition of anonymity, note that it can
be used to describe the anonymity provided by the Dining
Cryptographers protocol from \secref{protocols}. Recall that if one of
the cryptographers paid, the Dining Cryptographers protocol 
guarantees that each of the non-paying cryptographers think it possible
that any of the cryptographers but herself paid. In other words, if
$C=\{\mi{Alice},\mi{Betty},\mi{Charlene}\}$ are the cryptographers and
if cryptographer $i$ paid, then the protocol guarantees that
the paying action is anonymous up to $C\setminus\{j\}$ with respect to
cryptographer $j$, as long as $j\ne i$.

\paragraph{Coercion Resistance.}
Voting protocols are protocols in which anonymity plays an
important role.  
Voting protocols furthermore satisfy other interesting security
properties. 
Aside from secrecy of votes (that every voter's choice
should be private, and observers should not be able to figure out who
voted how), other properties include fairness (voters do not have any
knowledge of the distribution of votes until the final tallies are
announced), verifiability (every voter should be able to check whether
her vote was counted properly), and receipt freeness (no voter
has the means to prove to another that she has voted in a particular
manner). 

This last property, receipt freeness, is particularly interesting in
terms of epistemic content. Roughly speaking, receipt freeness says
that a voter Alice cannot prove to a potential coercer Corinna that
she voted in a particular way. This is the case even if Alice wishes
to cooperate with Corinna; receipt freeness guarantees that such
cooperation cannot lead to anything because it will be impossible for
Corinna to be certain how Alice voted. In that sense, receipt freeness
goes further than secrecy of votes. Even if Alice tells Corinna that
she voted a certain way, Corinna has no way to verify Alice's
assertion, and Alice has no way to convince her. 

Coercion resistance is closely related to receipt freeness but is
slightly stronger.
Intuitively, a voting protocol is coercion resistant if it prevents
voter coercion and vote buying even by active coercers: a coercer
should not be able to influence the behavior of a voter.  
Coercion resistance can be modeled epistemically, although the details
of the modeling is subtle, and many important details will be skipped
in the description below.
Part of the difficulty and subtlety is that the idea of coercion means
changing how a voter behaves based on a coercer's desired outcome or
goal, which needs to be modeled somehow.

One formalization of coercion resistance uses a model of voting
protocols based on traces, where some specific agents are highlighted:
a voter that the coercer tries to influence (called the coerced
voter), the coercer, and the remaining agents and authorities, assumed
to be honest.  
Every voter in the system votes according to a voting strategy, which
in the case of honest voters is the strategy corresponding to the
voting protocol.

The formalization assumes that every voter has a specific
voting goal, formally captured by the set of traces in which that
voter successfully votes according to her desired voting goal.
The coercer, however, is intent on affecting the coerced
voter---for instance, to coerce a vote for a given candidate, or
perhaps to coerce a vote away from a given candidate. 
To coerce a voter, the coercer hands the coerced voter a particular
strategy that will fulfill the coercer's goals instead of the coerced
voter's. For instance, the coercer's strategy may simply be one that
forwards all messages to and from the coercer, effectively making the
coerced voter a proxy for the coercer. 

Let 
$V$ be the space of possible strategies that voters and coercers can
follow. 
Coercion resistance can be defined by saying
that for every possible strategy $v\in V$, there is another strategy
$v'\in V$ that the coerced voter can use instead of $v$ with the
property that: (1) the voter always achieves her goal by using $v'$,
and (2) the coercer does not know whether the coerced voter used
strategy $v$ or $v'$.  
In other words, in every trace in which the coerced voter uses
strategy $v$, the coercer considers it possible, given her view of the trace, 
that the coerced voter is using strategy $v'$ instead. Conversely, in
every trace in which the coerced voter uses strategy $v'$, the coercer
considers it possible that the coerced voter is using strategy $v$. So,
the coercer cannot know whether the coerced voter followed the
coercer's instructions (i.e., used $v$) or tried to achieve her
own goal (i.e., used $v'$).  
As in the case of information flow in event systems in
\secref{events}, the definition of coercion resistance is a form of
closure property on traces, which corresponds to lack of knowledge in
the expected way, where knowledge is captured by an
indistinguishability relation on states based on the coercer's
observations.

\paragraph{Zero Knowledge.}
The property \emph{an agent does not learn anything about something},
as embodied in information-flow security policies and other forms of
confidentiality, is generally modeled using an indistinguishability
relation over states and enforced by making sure that there are enough
states to prevent the confidential information from being known by
unauthorized agents.

Another approach to modeling and enforcing this lack of learning is
demonstrated by \emph{zero knowledge interactive proof systems}. 
An interactive proof system for a string language $L$ is a two-party
system $(P,V)$ in which a prover $P$ tries to convince a verifier $V$
that some string $x$ is in $L$ through a sequence of message exchanges
amounting to an interactive proof of $x\in L$.
Classically, the prover is assumed to be infinitely powerful, while
the verifier is assumed to be a probabilistic polynomial-time Turing
machine.  
An interactive proof system has the property that if $x\in
L$, the conversation between $P$ and $V$ will show $x\in L$ with high
probability, and if $x\not\in L$, the conversation between \emph{any}
prover and $V$ will show $x\in L$ with low probability. (The details
for why the second condition refers to any prover rather than just $P$
is beyond the scope of this discussion.)

An interactive proof system for $L$ is zero knowledge if
whenever $x\in L$ holds
the verifier is able to generate \emph{on its own} the conversations
it would have had with the prover during an interactive proof of $x\in L$.
The intuition here is that the verifier does not learn anything from a
conversation with the prover (other than $x\in L$) if it can
learn exactly the same thing by generating that
whole conversation itself. 
Thus, the only knowledge gained by the verifier is that which the
prover initially set out to prove. 

Zero knowledge interactive proof systems rely on indistinguishability,
but not indistinguishability among a
large set of states. Rather, it is indistinguishability between two
scenarios: a scenario where the verifier interacts with the prover,
and a scenario where the verifier does not interact with the prover
but instead simulates a complete interaction with the prover. This
simulation paradigm, a core notion in modern theoretical computer
science, says roughly that an agent does not gain any knowledge
from interacting with the outside world if she can achieve the same
results without interacting with the outside world. 

To give a sense of the kinds of definitions that arise in this context,
here is one formal definition of perfect zero
knowledge: 
\begin{quote}
Let $(P,V)$ be an interactive proof system for $L$, where $P$ (the
prover) is an interactive Turing machine and $V$ (the verifier) is a
probabilistic polynomial-time interactive Turing machine. System
$(P,V)$ is \emph{perfect zero-knowledge} if for every probabilistic
polynomial-time interactive Turing machine $V^*$ there is a
probabilistic polynomial-time Turing machine $M^*$ (the simulator)
such that for every $x\in L$ the following two random variables are
identically distributed:
\begin{enumerate}
\item[(i)] the output of $V^*$ interacting with $P$
  on common input $x$; 
\item[(ii)] the output of machine $M^*$ on input $x$.
\end{enumerate}
\end{quote}
While the details are beyond the scope of this chapter, the intuition
behind this definition is to have, for every possible verifier $V^*$
(and not only $V$) interacting with $P$, a machine $M^*$ that can
simulate the interaction of $V^*$ and $P$ even though it does not have
access to the prover $P$. The existence of such simulators implies
that $V^*$ does not gain any knowledge from $P$. 

This gives a different epistemic foundation for confidentiality, one
that is intimately tied to computation and its complexity. The
relationship with classical epistemic logic is essentially
unexplored.


\section{Perspectives}
\label{chap13:perspectives}

The preceding sections illustrate how extensively epistemic concepts,
explicitly framed as an epistemic logic or not, have been
applied to security research. 
Whether the application of these concepts has been successful is a more
subjective question. 


In a certain sense, the problems described in this chapter are solved
problems by now. Confidentiality and authentication in cryptographic
protocol analysis under a formal model of cryptography and Dolev-Yao
attackers, for example, can be checked quite efficiently with a vast
array of methods, at least for common security properties, and the
definitions used approximate the epistemic definitions quite
closely.

So what are the remaining challenges in cryptographic protocol
analysis, and has epistemic logic a role to play?
The most challenging aspect of cryptographic protocol analysis is to
move beyond Dolev-Yao attackers and beyond formal models of
cryptography, towards more concrete models of cryptography.

Moving beyond a Dolev-Yao attacker requires shifting the notion of
message knowledge to use richer algebras of message with more
operations. Directions that have been explored include providing
attackers with the ability to perform offline dictionary attacks,
working with an XOR operation, or even number-based operations
such as exponentiation. 
One problem is that when the algebra of messages is subject to too
many algebraic properties, determining whether an attacker knows a
message may quickly become undecidable. 
Even when message knowledge for an attacker is decidable, it may
still be too complex for efficient reasoning. 
It is not entirely clear how epistemic concepts can help solve
problems in that arena. 


Moving from a formal model of cryptography to a
concrete model, one that reflects
real encryption schemes more accurately using sequences of bits and
computational indistinguishability, requires
completely shifting the approach to cryptographic protocol analysis. 

Formal models of cryptography work by abstracting away the
\emph{one-way security} property of encryption schemes---that it is
computationally hard to recover the sourcetext from a ciphertext
without knowing the encryption key. 
More concrete models of cryptography rely on stronger properties than
one-way security, properties such as \emph{semantic security},
which intuitively says that if any information about a message $m$ can
be computed by an efficient algorithm given the ciphertext $e_k(m)$
for a random $k$ and $m$ chosen according to an arbitrary probability
distribution, that same information can be computed without knowing
the ciphertext. In other words, the ciphertext $e_k(m)$ offers no
advantage in computing information about some message $m$ chosen from
an arbitrary probability distribution. 

The definition of semantic security is reminiscent of the
definition of zero knowledge interactive proof systems in
\secref{beyond}, and it is no accident, as they both rely on a
simulation paradigm to express the fact that no knowledge is gained. 
As in the case of zero knowledge interactive proof systems, there is a
clear epistemic component to the definition of semantic security, one
to which classical epistemic logic has not been applied. 

The main difficulty with applying classical epistemic logic to
concrete models of cryptography is that these models take attackers to
be probabilistic polynomial-time Turing machines, and take security
properties to be probabilistic properties relative to those
probabilistic polynomial-time Turing machines. This means that an
epistemic approach to concrete models of cryptography needs to 
be probabilistic as well as computationally  bounded. 
The former is not a problem, since probabilistic
reasoning shares much of the same foundations as epistemic reasoning. 
But the latter is more complicated. 
Concrete models of cryptography are not based on impossibility, but on
computational hardness.
And while possible-worlds definitions of knowledge are well suited to
talking about impossible versus possible outcomes, they fare 
less well at talking about difficult versus easy outcomes.

The trouble that possible-worlds definitions of knowledge run into
when trying to incorporate a notion of computational difficulty is
really the problem of logical omniscience in epistemic logic under a
different guise. Agents, under standard possible-worlds definitions of
knowledge, know all tautologies, and know all logical consequences of
their knowledge: if $K\phi$ is true and $\phi\rimp\psi$ is valid, then
$K\psi$ is also true. Any normal epistemic operator will satisfy these
properties, and in particular, any epistemic logic based on Kripke
structures will satisfy these properties. Normality does not deal well
with computational difficulty, because while it may be computationally
difficult to establish that $\phi\rimp\psi$ is valid, a normal modal
logic will happily derive all knowledge-based consequences of that
valid formula. 
It would seem that giving a satisfactory epistemic account of concrete
models of cryptography requires a non-normal epistemic logic, one that
supports a form of resource-bounded knowledge. 
Resource-bounded knowledge is not well understood, and logics
for resource-bounded knowledge still feel too immature to form a
solid basis for reasoning about concrete models of cryptography.


Leaving aside concrete models of cryptography, it is almost impossible
to discuss epistemic logic in the context of 
cryptographic protocols without addressing the issue of BAN Logic. 
BAN Logic is an interesting and original use of logic, developed to
prove cryptographic protocol
properties manually by paralleling informal arguments for protocol
correctness. 

BAN Logic has spilled a lot of virtual ink. Aside from its technical
limitations---it requires a protocol idealization step that remains
outside the purview of the logic but affects the results of
analysis---the logic is considered somewhat \emph{pass\'{e}}. 
Other approaches we saw in \secref{reasoning} operate in the same space,
namely analyzing 
cryptographic protocols under a formal model of cryptography in the
presence of Dolev-Yao attackers, and most are less limited
and more easily automated.
Other approaches, such as Protocol Composition Logic, even advocate
Hoare-style reasoning about the protocol text from within the logic,
just like BAN Logic.

My perspective on BAN Logic is that it tried to do something which has
not really been tried since, something that remains a sort of litmus
test for our understanding of security in cryptographic protocols:
identifying high-level primitives that capture relevant concepts for
security, high-level primitives that match our intuitive understanding
of security properties, those same intuitions that guide our design of
cryptographic protocols in the first place. 
We do not have such high-level primitives in any other
framework, all of which tend to work at much lower levels of
abstraction.
The primitives in BAN Logic are intuitively attractive, but poorly
understood. 
The continuing conversation on BAN Logic is a reminder that we still
do not completely understand the basic concepts and basic terms needed
to discuss cryptographic protocols, and I think BAN Logic remains
relevant, if only as a nagging voice telling us that we have
not quite gotten it right yet.


Many of the issues that arise when trying to push cryptographic
protocol analysis from a formal model of cryptography to a more
concrete model also come up in the context of information-flow
security. As mentioned in \secref{languages}, recent work has turned
to the question of declassification, or controlled release of
information. The reason for this is purely pragmatic: most
applications need to release some kind of information in order to do
any useful work, even under a lax interpretation of noninterference.  

But it does not take long to see that even a controlled release of
information can lead to unwanted release of information in the
aggregate. Returning to the password-login problem from
\secref{languages}, it is clear that every wrong attempt at entering a
password leaks some information, something that needs to happen if the
login screen is to operate properly. 
But of course, repeated attempts at checking the password will
eventually lead to the correct password as the only remaining possibility,
which is a severe undesirable release of information. 
Security policies controlling declassification therefore seem to
require a way to account for more quantitative notions of leakage
which aggregate over time, something that symbolic approaches
to information-flow security have difficulty handling well. 

Modeling information flow quantitatively can be seen as a move from
reasoning about information as a monolithic unit to reasoning about
information as a resource. Once we make that leap, other resources
affecting information flow start suggesting themselves. 
For example, execution time can leak information. Consider the simple
program: 
\begin{align*}
 & \ms{if}~\text{\textit{(high-security Boolean variable)}}\\
 & \qquad\ms{then}~\text{\textit{fast code}}\\
 & \qquad\ms{else}~\text{\textit{slow code}}
\end{align*}
By observing the execution time of the program, we can determine the
value of the high-security Boolean variable. This example is rather
silly, of course, but it illustrates the point that information
leakage can occur based on observations of other resources than simply
the state of memory. 

What about the combination of information flow and cryptography?
After all, in practice, systems do use cryptography internally to help
keep data confidential.
Encrypted data can presumably be written on shared storage (which 
might be easier to manage than storage segregated into high-security
and low-security storage) or moved online, or
in general given to low-security agents without information being
released, as
long as they do not have the key or the resources to decrypt.
Accounting for cryptography in information-flow security 
raises questions similar to those in cryptographic protocol analysis
concerning what models of cryptography to use and how to account for
the cryptographic capabilities of attackers. It also raises
difficulties similar to those in cryptographic protocol analysis when
trying to move from a formal model of cryptography to a concrete
model, including how to provide an epistemic foundation for information
flow using a resource-bounded definition of knowledge.

\paragraph{Conclusion.} 
Epistemic concepts are central to many aspects of reasoning about
security. 
In some cases, these epistemic concepts may even naturally take
expression in a \emph{bona fide} epistemic logic that can be used to
formalize the reasoning. But whether an epistemic logic is used or
not, the underlying concepts are clearly epistemic. In particular, the
notion of truth at all possible worlds reappears in many different
guises throughout the literature. 

Research in security analysis has reached a sort of convergence
point around the use of symbolic methods.
The challenge seems to be to move beyond this convergence point, and
such a move requires taking resources seriously:
realistic definitions of security rely on the notion that exploiting a
vulnerability should require more resources (time, power, information)
than are realistically available to an attacker. In epistemic terms,
what is needed is a reasonably well-behaved definition of
resource-bounded knowledge, itself an active area of
research in epistemic logic. It would appear, then, that advances in
epistemic logic may well help increase our ability to reason about
security in direct ways.

\paragraph{Acknowledgments.} Thanks to Aslan Askarov, Philippe
Balbiani, Stephen Chong, and Vicky Weissman for comments on an early
draft of this chapter.


\section{Bibliographic Notes and Further Reading}
\label{chap13:bibnotes}

For the basics of epistemic logic, both the syntax and the semantics,
the reader is referred to the introductory chapter of the current
volume. For the sake of making this chapter as self-contained as
possible, most of the background material can be usefully obtained
from the textbooks of Fagin, Halpern, Moses, and Vardi
\cite{r:fagin95} and Meyer and Van der Hoek \cite{r:meyer95}. The
possible-worlds definition of knowledge used throughout this
chapter is simply the view that knowledge is truth at all worlds that
an agent considers as possible alternatives to the current world, a
view which goes back to Hintikka~\cite{r:hintikka62}.


\paragraph{Cryptographic Protocols.}

While the focus of the section is on symbolic cryptographic protocol
analysis, cryptographic protocols can also be studied from the
perspective of more computationally-driven cryptography, of the kind
described in \secref{perspectives}; see Goldreich \cite{r:goldreich04}.
The Russian Cards problem, which was first presented at the Moscow
Mathematic Olympiads in 2000, is described formally and studied from
an epistemic perspective by Van Ditmarsch \cite{r:ditmarsch03}. 
The problem has been used as a benchmark for several
epistemic logic model checkers \cite{r:ditmarsch06}. 
The Dining Cryptographers problem and the corresponding protocol is
described by Chaum \cite{r:chaum88}. It was proved correct in an
epistemic temporal logic model checker by Van der Meyden and Su
\cite{r:meyden04}.


For a good overview of classical cryptography along with some
perspectives on protocols, see Stinson \cite{r:stinson95} and
Schneier \cite{r:schneier96}; both volumes contain descriptions of
DES, AES, RSA, and elliptic-curve cryptography.
Goldreich \cite{r:goldreich01,r:goldreich04} is also introductory, but
from the perspective of modern computational cryptography.


For a good high-level survey of the kind of problems surrounding the design and
deployment of cryptographic protocols, see Anderson
and Needham \cite{r:anderson95}, then follow up with Abadi and
Needham's \cite{r:abadi96a} prudent engineering practices. 
The key distribution protocol used as the first example in
\secref{crypto} is related to the Yahalom protocol described
by Burrows, Abadi, and Needham \cite{r:burrows90}. The
Needham-Schroeder protocol was first described in Needham and
Schroeder \cite{r:needham78}. 
The man-in-the-middle attack on the Needham-Schroeder protocol in the
presence of an insider attacker was pointed out by Lowe
\cite{r:lowe95}, and the fix was analyzed by Lowe \cite{r:lowe96}.


The Dolev-Yao model of the attacker given in \secref{attackers} is
due to Dolev and Yao \cite{r:dolev83}. 


The formal definition of message knowledge via $\mi{Analyzed}$
and $\mi{Synthesized}$ sets is taken from Paulson
\cite{r:paulson98}. Equivalent definitions are given in nearly every
formal system for reasoning about cryptographic protocols in a formal
model of cryptography.  
Message knowledge can be defined using a local deductive system,
which makes it fit nicely within the deductive knowledge framework of Konolige
\cite{r:konolige86}---see also Pucella \cite{r:pucella06c}. 
More generally, message knowledge is a form of
algorithmic knowledge \cite{r:halpern94}, that is, a local form
of knowledge that relies on an algorithm to compute what an agent knows
based on the local state of the agent. In the case of a Dolev-Yao
attacker, this local algorithm simply computes the sets of analyzed
and synthesized messages \cite{r:halpern12}.  

Another way of defining message knowledge is the hidden automorphism
model, due to Merritt \cite{r:merritt83}, which is a form of
possible-worlds knowledge. While it never gained much
traction, it has been used in later work by Toussaint and Wolper
\cite{r:toussaint89} and also in the logic of Bieber
\cite{r:bieber90}.
It uses algebraic presentations of encryption
schemes called cryptoalgebras. 
There is a unique surjective cryptoalgebra homomorphism from the free
cryptoalgebra over a set of plaintexts and keys to any cryptoalgebra
over the same plaintexts and keys which acts as the identity on plaintexts
and keys. 
Message knowledge in a given cryptoalgebra $C$ is knowledge of the
structure of messages as given by that surjective homomorphism from
the free cryptoalgebra to $C$.
A revealed reduct is a subset of $C$ that the agent has seen.
A state of knowledge with respect to revealed reduct $R$ is a set of
of mappings $f$ from the free cryptoalgebra to $C$ that are
homomorphisms on $f^{-1}(R)$. 
In this context, an agent knows message $m$ if the agent knows the
representation of message $m$, meaning that $m$ is the image of the same
free cryptoalgebra term under every mapping in the state of knowledge
of the agent.
Thus, if an agent receives $\sencr{m_1}{k}$ and $\sencr{m_2}{k}$
but does not receive $k$, then only $\sencr{m_1}{k}$ and
$\sencr{m_2}{k}$ are in the revealed reduct; the agent may consider
any distinct messages $m'_1$ and $m'_2$ to map to $\sencr{m_1}{k}$ and
$\sencr{m_2}{k}$ after encryption with $k$, since any such mapping
will act as a homomorphism on the pre-image of the revealed reduct.

Possible-worlds definitions of knowledge in the presence of
cryptography are problematic because cryptography affects what agents
can observe, and this impacts the definition of the accessibility
relation between worlds. 
The idea of replacing encrypted messages in the local state of agents
by a token goes back to Abadi and Tuttle's semantics for BAN Logic
\cite{r:abadi91}. 
Treating encrypted messages as tokens while still allowing agents to
distinguish different encrypted messages is less common, but has been
used at least by Askarov and Sabelfeld \cite{r:askarov07} and Askarov,
Hedin, and Sabelfeld \cite{r:askarov08} in the context of information
flow.


There are several frameworks for formally reasoning about
cryptographic protocols, and I shall not list them all here. But I
hope to provide enough pointers to the literature to ensure that 
the important ones are covered.
For an early survey on the state of the art in formal reasoning about
cryptographic protocols until 1995, see Meadows
\cite{r:meadows95}.


The Inductive Method described in \secref{reasoning} is
due to Paulson \cite{r:paulson98}, and is built atop the Isabelle logical
framework \cite{r:paulson94}, a framework for higher-order logic. 
BAN Logic is introduced by Burrows, Abadi, and Needham
\cite{r:burrows90}, who use it to perform an analysis of several
existing protocols in the literature. The logic courted controversy
pretty much right from the start \cite{r:nessett90,r:burrows90a}. 
Probably the most talked-about problem with BAN Logic is the lack of
an independently-motivated semantics which would ensure that 
statements of the logic match operational intuition. Without such a
semantics, it is difficult to argue for the reasonableness of the
result of a BAN Logic analysis, except for the pragmatic observation that 
failure to prove a statement in BAN Logic often indicates a problem
with the cryptographic protocol.  
Abadi and Tuttle
\cite{r:abadi91} attempt to remedy the situation by defining a
semantics for BAN Logic. 
Successor logics
extending or modifying BAN Logic generally start with a variant of
the Abadi-Tuttle semantics
\cite{r:syverson90,r:gong90,r:oorschot93a,r:syverson94,r:wedel96,r:stubblebine96}. 
Contemporary epistemic logic alternatives to BAN Logic were also developed,
using a semantics in terms of protocol execution, but they
never really took hold \cite{r:bieber90,r:moser90}.


The model checker MCK is described by Gammie and Van der Meyden
\cite{r:gammie04}, and was used to analyze the Dining Cryptographers
protocol \cite{r:meyden04} as well as the Seven Hands protocol for the
Russian Cards problem \cite{r:ditmarsch06}. 
TDL is an alternative epistemic temporal logic for reasoning
about cryptographic protocols with a model checker developed by Penczek and
Lomuscio \cite{r:lomuscio06}, based on a earlier model checker
\cite{r:raimondi04}. 
TDL is a branching-time temporal epistemic logic extended with a
message knowledge primitive in addition to standard possible-worlds
knowledge for expressing higher-order knowledge, and does not provide
explicit  support for attackers in its modeling language. 
The model-checking complexity results mentioned are due to Van der
Meyden and Shilov \cite{r:meyden99}; see also 
Engelhardt, Gammie, and Van der Meyden \cite{r:meyden07} and Huang and
Van der Meyden \cite{r:meyden10}.

Another epistemic logic which forms the basis for reasoning about
cryptographic protocol is Dynamic Epistemic Logic (DEL)
\cite{r:gerbrandy99}. 
DEL is an epistemic logic of broadcast announcements which includes formulas of the form $[\rho]_i\phi$, read
\emph{$\phi$ holds after agent $i$ broadcasts formula $\rho$}, where $\rho$
is a formula in a propositional epistemic sublanguage. (The actual syntax of DEL
is slightly different.)
Agents may broadcast that they know a fact, and this broadcast affects
the knowledge of other agents.
Kripke structures are used to capture the state of knowledge of agents
at a point in time, and agent $i$ announcing $\rho$ will change 
Kripke structure $M$ representing the current state of knowledge of
all agents into a 
Kripke structure $M^{\rho,i}$ representing the new state of
knowledge that obtains. 
Dynamic Epistemic Logic has been used to analyze the Seven Hands
protocol in great detail \cite{r:ditmarsch03}. 
Extensions to handle cryptography are described by Hommersom, Meyer, and
De Vink \cite{r:hommersom04}, as well as Van Eijck and Orzan
\cite{r:vaneijck07}. 

Process calculi, starting with the spi calculus \cite{r:gordon99} and
later the applied pi calculus \cite{r:abadi01}, have been particularly
successful tools for reasoning about cryptographic protocols. These
use either observational equivalence to show that a process
implementation of the protocol is equivalent to another process that
clearly satisfies the required properties, or static analysis
such as type checking to check the properties \cite{r:gordon03}.
Epistemic logics defined against models obtained from processes are
given by Chadha, Delaune, and Kremer \cite{r:chadha09} and Toninho and
Caires \cite{r:toninho09}.
Another process calculus, CSP, has also proved popular as a foundation
for reasoning about cryptographic protocols \cite{r:lowe98,r:ryan00}.

Finally, other approaches rely on logic programming ideas: the NRL
protocol analyzer \cite{r:meadows96}, Multiset Rewriting
\cite{r:cervesato99}, and ProVerif \cite{r:blanchet01}.  
Thayer, Herzog, and Guttman \cite{r:thayer99} introduce a distinct
semantic model for protocols, strand spaces, which has some advantages
over traces. Syverson \cite{r:syverson99} develops an
authentication logic on top of strand spaces, while Halpern and
Pucella \cite{r:halpern03d} investigate the suitability of strand
spaces as a basis for epistemic reasoning.


\paragraph{Information Flow.}

Bell and LaPadula \cite{r:bell73a,r:bell73b} were among the first to develop
mandatory access control, and introducing the idea of attaching
security levels to data to enforce confidentiality. 

Early work on information flow security mostly focused on event
traces, and tried to describe both closure conditions on traces, as
well as unwinding conditions that would allow one to check that a set
of event traces satisfies the security condition. Separability was
defined by McLean \cite{r:mclean94}, noninference by O'Halloran
\cite{r:oHalloran90}, generalized noninference by McLean
\cite{r:mclean94}, and generalized noninterference by McCullough
\cite{r:mccullough87,r:mccullough88} following the work of Goguen and
Meseguer \cite{r:goguen82,r:goguen84}. Other definitions of
information-flow security for event systems are given by Sutherland
\cite{r:sutherland86} and Wittbold and Johnson \cite{r:wittbold90}. 
A modern approach to information-flow security in event systems is
described by Mantel \cite{r:mantel00}.
The set-theoretic description of the knowledge operator is taken from
Halpern \cite{r:halpern99}, but appears in various guises in the
economics literature~\cite{r:aumann89}. 
Halpern and O'Neill \cite{r:halpern08} layer an explicit epistemic
language on top of the event models re-expressed as Kripke structures,
and show that the resulting logic can capture common definitions of
confidentiality in event systems.


Denning and Denning \cite{r:denning77} first pointed out that
programming languages are a useful setting for reasoning about
information flow by observing that static analysis can be used to
identify and control information flow. 
Most recent work on information-flow security from a programming
language perspective goes back to Heintze
and Riecke's Secure Lambda Calculus \cite{r:heintze98} in a
functional language setting, and Smith and Volpano
\cite{r:smith98} in an imperative language setting. Honda, Vasconcelos, and
Yoshida \cite{r:honda00} give a similar development in the context of
a process calculus. Sabelfeld and Myers
\cite{r:sabelfeld03} give a survey and overview of the state of the
field up to 2003.
Balliu, Dam, and Le Guernic \cite{r:balliu11,r:balliu12} offer a rare
use of an explicit epistemic temporal logic to reason about
information-flow security.  
Sabelfeld and Sands
\cite{r:sabelfeld09} give a good overview of the issues involved in declassification for language-based information flow.
Askarov and Sabelfeld \cite{r:askarov07} use an epistemic
logic in the context of declassification. 
Chong~\cite{r:chong10} uses a form of algorithmic knowledge to model
information release requirements.


\paragraph{Beyond Confidentiality.}

Protocols for anonymous communication generally rely on a cloud of
intermediaries that prevent information about the identity of the
original sender to be isolated; Crowds is an example of such a
protocol~\cite{r:reiter98}. Anonymity has been well studied as an
instance of confidentiality \cite{r:hughes04,r:garcia05}. The
explicit connection with epistemic logic was made by Halpern and
O'Neill \cite{r:halpern05d}, which is the source of the definitions in
\secref{beyond}. An early analysis of anonymity via epistemic logic
is given by Syverson and Stubblebine \cite{r:syverson99a}. 

Anonymity is an important component of voting protocols. Van Eijck and
Orzan \cite{r:vaneijck07} prove anonymity for a specific voting
protocol using epistemic logic.
More general analyses of voting protocols with epistemic logic have
also been attempted \cite{r:baskar07,r:kusters09}. 
The model of coercion resistance in \secref{beyond} is from 
K\"usters and Truderung \cite{r:kusters09}.

Zero knowledge interactive proof systems were introduced by
Goldwasser, Micali, and Rackoff \cite{r:goldwasser89} and
have become an important tool in theoretical computer science.
A good overview is given by Goldreich \cite{r:goldreich01}.
Halpern, Moses, and Tuttle \cite{r:halpern88} give an epistemically-motivated
analysis of zero knowledge interactive proof systems using a
computationally-bounded definition of knowledge devised by Moses 
\cite{r:moses88}. 

Another context in which epistemic concepts---or perhaps more
accurately, epistemic vocabulary---appear is that of authorization and
trust management. Credential-based authorization policies can be used
to control access to resources by requiring agents to present
appropriate credentials (such as certificates) proving that they are
allowed access. 
Because systems that rely on credential-based
authorization policies are often decentralized systems, meaning that
there is no central clearinghouse for determining for every
authorization request whether an agent has the appropriate
credentials, the entire approach relies on a web of trust between agents
and credentials. Since in many such systems credentials can be
delegated---an agent may allow another agent to act on her
behalf---not only can credential checking become complicated, but
authorization policies themselves become nontrivial to analyze to
determine contradictions (an action being both allowed and forbidden
by the policy under certain conditions) or coverage (a class
of actions remaining unregulated by the policy under certain
conditions). 
Where do epistemic concepts come up in such a scenario? Authorization
logics from the one introduced by Abadi, Burrows, Lampson, and Plotkin
\cite{r:abadi93} to the recent NAL \cite{r:schneider11} have been
described as logics of belief, and are somewhat reminiscent of BAN
Logic. One of their basic primitives is a formula $A~
\mathsf{says}~F$, which as a credential means that $A$ believes and is
accountable for the truth of $F$. Delegation, for example, is captured
by a formula $(A~\mathsf{says}~F) \rimp (B~\mathsf{says}~G)$. This
form of belief is entirely axiomatic, just like belief in BAN Logic.


\paragraph{Perspectives.}

Ryan and Schneider \cite{r:ryan98} have extended the Dolev-Yao model
of attackers with an XOR operation; Millen and Shmatikov
\cite{r:millen03} with products and enough exponentiation to model the
Diffie-Hellman key-establishment protocol~\cite{r:diffie76}; and Lowe
\cite{r:lowe02} and later Corin, Doumen, and Etalle
\cite{r:corin05} and Baudet \cite{r:baudet05} with the ability to
mount offline dictionary attacks. As described by Halpern and Pucella
\cite{r:halpern12}, many of these can be expressed using algorithmic
knowledge, at least in the context of eavesdropping attackers.
More generally, extending Dolev-Yao with additional operations can
best be studied using equational theories, that is, 
equations induced by looking at the algebra of the additional
operations; see for example Abadi and Cortier \cite{r:abadi04} and
Chevalier and Rusinowitch \cite{r:chevalier08}.

While it would be distracting to discuss the back and forth over BAN
Logic in the decades since its inception, I will point out that recent
work by Cohen and Dam has taken a serious look at the logic with
modern eyes, and highlighted both interesting interpretations as well
as subtleties \cite{r:cohen05,r:cohen05a}.
The protocol composition logic PCL of Datta, Derek, Mitchell, and Roy
\cite{r:datta07}, which builds on earlier work by Durgin, Mitchell,
and Pavlovic \cite{r:durgin03}, is a modern attempt at devising a
logic for Hoare-style reasoning about cryptographic protocols.

A good overview of concrete models of cryptography is given by
Goldreich \cite{r:goldreich01}. Semantic security, among others, is
studied by Bellare, Chor, Goldreich, and Schnorr \cite{r:bellare98}.
The relationship between formal models of
cryptography and concrete models---how well does the former 
approximate the latter?---has been explored by Abadi and Rogaway
\cite{r:abadi02a}, and later extended by Micciancio and Warinschi
\cite{r:micciancio04}, among others. Backes, Hofheinz, and Unruh
\cite{r:backes09} provide a good overview.

Approaches to analyze cryptographic protocols in a concrete model
of cryptography have been developed \cite{r:lincoln98, r:mitchell01}. 
In recent years some of the
approaches for analyzing cryptographic protocols in a formal model of
cryptography have been modified to work with a concrete
model of cryptography, such as PCL~\cite{r:datta05} and ProVerif \cite{r:blanchet08}. 
In some cases, cryptographic protocol analysis in a concrete model
relies on extending indistinguishability over states
to indistinguishability over the whole protocol
\cite{r:datta04}.  

Defining a notion of resource-bounded knowledge that does not suffer
from the logical omniscience problem is an ongoing research project in
the epistemic logic community, and various approaches have been
advocated, each with its advantages and its deficiencies: algorithmic
knowledge \cite{r:halpern94}, impossible possible worlds \cite{r:hintikka75},
awareness \cite{r:fagin88}. A comparison between the approaches in terms of
expressiveness and pragmatics appears in Halpern and
Pucella~\cite{r:halpern11}.

Information flow in probabilistic programs was first
investigated by Gray and Syverson \cite{r:gray98} using probabilistic
multiagent systems \cite{r:halpern93}. Backes and Pfitzmann
\cite{r:backes02} study it in a more computational setting. 
Smith \cite{r:smith09} presents some of the tools that need to be
considered to analyze the kind of partial information leakage
occurring in the password-checking example.
Preliminary work on information flow in the presence of cryptography
includes Laud \cite{r:laud03}, Hutter and Schairer
\cite{r:hutter04}, and Askarov, Hedin, and Sabelfeld
\cite{r:askarov08}.


\bibliographystyle{plain} 
{\small \bibliography{knowledgeandsecurity}}  

\begin{thebibliography}{100}

\bibitem{r:abadi93}
M.~Abadi, M.~Burrows, B.~Lampson, and G.~Plotkin.
\newblock A calculus for access control in distributed systems.
\newblock {\em ACM Transactions on Programming Languages and Systems},
  15(4):706--734, 1993.

\bibitem{r:abadi04}
M.~Abadi and V.~Cortier.
\newblock Deciding knowledge in security protocols under equational theories.
\newblock In {\em Proc.~31st Colloquium on Automata, Languages, and Programming
  (ICALP'04)}, volume 3142 of {\em Lecture Notes in Computer Science}, 2004.

\bibitem{r:abadi01}
M.~Abadi and C.~Fournet.
\newblock Mobile values, new names, and secure communication.
\newblock In {\em Proc.~28th Annual ACM Symposium on Principles of Programming
  Languages (POPL'01)}, pages 104--115, 2001.

\bibitem{r:gordon99}
M.~Abadi and A.~D. Gordon.
\newblock A calculus for cryptographic protocols: The spi calculus.
\newblock {\em Information and Computation}, 148(1):1--70, 1999.

\bibitem{r:abadi96a}
M.~Abadi and R.~Needham.
\newblock Prudent engineering practice for cryptographic protocols.
\newblock {\em IEEE Transactions on Software Engineering}, 22(1):6--15, 1996.

\bibitem{r:abadi02a}
M.~Abadi and P.~Rogaway.
\newblock Reconciling two views of cryptography (the computational soundness of
  formal encryption).
\newblock {\em Journal of Cryptology}, 15(2):103--127, 2002.

\bibitem{r:abadi91}
M.~Abadi and M.~R. Tuttle.
\newblock A semantics for a logic of authentication.
\newblock In {\em Proc.~10th ACM Symposium on Principles of Distributed
  Computing (PODC'91)}, pages 201--216, 1991.

\bibitem{r:anderson95}
R.~Anderson and R.~Needham.
\newblock Programming {Satan}'s computer.
\newblock In J.~van Leeuwen, editor, {\em Computer Science Today: Recent Trends
  and Developments}, volume 1000 of {\em Lecture Notes in Computer Science},
  pages 426--440. Springer-Verlag, 1995.

\bibitem{r:askarov08}
A.~Askarov, D.~Hedin, and A.~Sabelfeld.
\newblock Cryptographically-masked flows.
\newblock {\em Theoretical Computer Science}, 402(2--3):82--101, 2008.

\bibitem{r:askarov07}
A.~Askarov and A.~Sabelfeld.
\newblock Gradual release: Unifying declassification, encryption and key
  release policies.
\newblock In {\em Proc.~2007 IEEE Symposium on Security and Privacy}, pages
  207--221. IEEE Computer Society Press, 2007.

\bibitem{r:aumann89}
R.~J. Aumann.
\newblock Notes on interactive epistemology.
\newblock Cowles Foundation for Research in Economics working paper, 1989.

\bibitem{r:backes09}
M.~Backes, D.~Hofheinz, and D.~Unruh.
\newblock {CoSP}: A general framework for computational soundness proofs.
\newblock In {\em Proc.~16th ACM Conference on Computer and Communications
  Security (CCS'09)}, pages 66--78. ACM Press, 2009.

\bibitem{r:backes02}
M.~Backes and B.~Pfitzmann.
\newblock Computational probabilistic non-interference.
\newblock In {\em Proc.~7th European Symposium on Research in Computer Security
  (ESORICS'02)}, volume 2502 of {\em Lecture Notes in Computer Science}, pages
  1--23. Springer-Verlag, 2002.

\bibitem{r:balliu11}
M.~Balliu, M.~Dam, and G.~{Le Guernic}.
\newblock Epistemic temporal logic for information flow security.
\newblock In {\em Proc.~ACM SIGPLAN 6th Workshop on Programming Languages and
  Analysis for Security (PLAS'11)}. ACM Press, 2011.

\bibitem{r:balliu12}
M.~Balliu, M.~Dam, and G.~{Le Guernic}.
\newblock {ENCOVER}: Symbolic exploration for information flow security.
\newblock In {\em Proc.~25th IEEE Computer Security Foundations Symposium
  (CSF'12)}, pages 30--44. IEEE Computer Society Press, 2012.

\bibitem{r:baskar07}
A.~Baskar, R.~Ramanujam, and S.~P. Suresh.
\newblock Knowledge-based modelling of voting protocols.
\newblock In {\em Proc.~11th Conference on Theoretical Aspects of Rationality
  and Knowledge (TARK'07)}, pages 62--71. ACM Press, 2007.

\bibitem{r:baudet05}
M.~Baudet.
\newblock Deciding security of protocols against off-line guessing attacks.
\newblock In {\em Proc.~12th ACM Conference on Computer and Communications
  Security (CCS'05)}, pages 16--25. ACM Press, 2005.

\bibitem{r:bell73a}
D.~E. Bell and L.~J. LaPadula.
\newblock Secure computer systems: Mathematical foundations.
\newblock Technical Report MTR-2547, Volume 1, MITRE Corporation, 1973.

\bibitem{r:bellare98}
M.~Bellare, B.~Chor, O.~Goldreich, and C.~Schnorr.
\newblock Relations among notions of security for public-key encryption
  schemes.
\newblock In {\em Proc.~18th Annual International Cryptology Conference
  (CRYPTO'98)}, volume 1462 of {\em Lecture Notes in Computer Science}, pages
  26--45. Springer-Verlag, 1998.

\bibitem{r:bieber90}
P.~Bieber.
\newblock A logic of communication in hostile environment.
\newblock In {\em Proc.~3rd IEEE Computer Security Foundations Workshop
  (CSFW'90)}, pages 14--22. IEEE Computer Society Press, 1990.

\bibitem{r:blanchet01}
B.~Blanchet.
\newblock An efficient cryptographic protocol verifier based on {Prolog} rules.
\newblock In {\em Proc.~14th IEEE Computer Security Foundations Workshop
  (CSFW'01)}, pages 82--96. IEEE Computer Society Press, 2001.

\bibitem{r:blanchet08}
B.~Blanchet.
\newblock A computationally sound mechanized prover for security protocols.
\newblock {\em IEEE Transactions on Dependable and Secure Computing},
  5(4):193--207, 2008.

\bibitem{r:burrows90}
M.~Burrows, M.~Abadi, and R.~Needham.
\newblock A logic of authentication.
\newblock {\em ACM Transactions on Computer Systems}, 8(1):18--36, 1990.

\bibitem{r:burrows90a}
M.~Burrows, M.~Abadi, and R.~Needham.
\newblock Rejoinder to {Nessett}.
\newblock {\em ACM Operating Systems Review}, 24(2):39--40, 1990.

\bibitem{r:cervesato99}
I.~Cervesato, N.~Durgin, P.~Lincoln, J.~Mitchell, and A.~Scedrov.
\newblock A meta-notation for protocol analysis.
\newblock In {\em Proc.~12th IEEE Computer Security Foundations Workshop
  (CSFW'99)}, pages 55--69. IEEE Computer Society Press, 1999.

\bibitem{r:chadha09}
R.~Chadha, S.~Delaune, and S.~Kremer.
\newblock Epistemic logic for the applied pi calculus.
\newblock In {\em Proc.~IFIP International Conference on Formal Techniques for
  Distributed Systems}, volume 5522 of {\em Lecture Notes in Computer Science},
  pages 182--197. Springer-Verlag, 2009.

\bibitem{r:chaum88}
D.~Chaum.
\newblock The dining cryptographers problem: Unconditional sender and recipient
  untraceability.
\newblock {\em Journal of Cryptology}, 1(1):65--75, 1988.

\bibitem{r:chevalier08}
Y.~Chevalier and M.~Rusinowitch.
\newblock Hierarchical combination of intruder theories.
\newblock {\em Information and Computation}, 206(2--4):352--377, 2008.

\bibitem{r:chong10}
S.~Chong.
\newblock Required information release.
\newblock In {\em Proc.~23rd IEEE Computer Security Foundations Symposium
  (CSF'10)}, pages 215--227. IEEE Computer Society Press, 2010.

\bibitem{r:cohen05a}
M.~Cohen and M.~Dam.
\newblock A completeness result for {BAN} logics.
\newblock In {\em Proc.~Methods for Modalities (M4M)}, pages 202--219, 2005.

\bibitem{r:cohen05}
M.~Cohen and M.~Dam.
\newblock Logical omniscience in the semantics of {BAN} logic.
\newblock In {\em Proc.~Workshop on Foundations of Computer Security (FCS'05)},
  pages 121--132, 2005.

\bibitem{r:corin05}
R.~Corin, J.~Doumen, and S.~Etalle.
\newblock Analysing password protocol security against off-line dictionary
  attacks.
\newblock In {\em Proc.~2nd International Workshop on Security Issues with
  Petri Nets and other Computational Models (WISP'04)}, volume 121 of {\em
  Electronic Notes in Theoretical Computer Science}, pages 47--63. Elsevier
  Science Publishers, 2005.

\bibitem{r:datta07}
A.~Datta, A.~Derek, J.~C. Mitchell, and A.~Roy.
\newblock {Protocol} {Composition} {Logic} ({PCL}).
\newblock {\em Electronic Notes in Theoretical Computer Science},
  172(1):311--358, 2007.

\bibitem{r:datta05}
A.~Datta, A.~Derek, J.~C. Mitchell, V.~Shmatikov, and M.~Turuani.
\newblock Probabilistic polynomial-time semantics for a protocol security
  logic.
\newblock In {\em Proc.~32nd Colloquium on Automata, Languages, and Programming
  (ICALP'05)}, pages 16--29, 2005.

\bibitem{r:datta04}
A.~Datta, R.~K{\"u}sters, J.~C. Mitchell, A.~Ramanathan, and V.~Shmatikov.
\newblock Unifying equivalence-based definitions of protocol security.
\newblock In {\em Proc.~Workshop on Issues in the Theory of Security
  (WITS'04)}, 2004.

\bibitem{r:denning77}
D.~E. Denning and P.~J. Denning.
\newblock Certification of programs for secure information flow.
\newblock {\em Communications of the ACM}, 20(7):504--513, 1977.

\bibitem{r:diffie76}
W.~Diffie and M.~E. Hellman.
\newblock New directions in cryptography.
\newblock {\em IEEE Transactions on Information Theory}, 22:664--654, 1976.

\bibitem{r:dolev83}
D.~Dolev and A.~C. Yao.
\newblock On the security of public key protocols.
\newblock {\em IEEE Transactions on Information Theory}, 29(2):198--208, 1983.

\bibitem{r:durgin03}
N.~A. Durgin, J.~C. Mitchell, and D.~Pavlovic.
\newblock A compositional logic for proving security properties of protocols.
\newblock {\em Journal of Computer Security}, 11(4):677--722, 2003.

\bibitem{r:meyden07}
K.~Engelhardt, P.~Gammie, and R.~van~der Meyden.
\newblock Model checking knowledge and linear time: {PSPACE} cases.
\newblock In {\em Proc.~Symposium on Logical Foundations of Computer Science},
  volume 4514 of {\em Lecture Notes in Computer Science}, pages 195--211.
  Springer-Verlag, 2007.

\bibitem{r:fagin88}
R.~Fagin and J.~Y. Halpern.
\newblock Belief, awareness, and limited reasoning.
\newblock {\em Artificial Intelligence}, 34:39--76, 1988.

\bibitem{r:fagin95}
R.~Fagin, J.~Y. Halpern, Y.~Moses, and M.~Y. Vardi.
\newblock {\em Reasoning about Knowledge}.
\newblock MIT Press, 1995.

\bibitem{r:gammie04}
P.~Gammie and R.~van~der Meyden.
\newblock {MCK}: Model checking the logic of knowledge.
\newblock In {\em Proc.~16th International Conference on Computer Aided
  Verification (CAV'04)}, Lecture Notes in Computer Science, pages 479--483.
  Springer-Verlag, 2004.

\bibitem{r:garcia05}
F.~D. Garcia, I.~Hasuo, W.~Pieters, and P.~van Rossum.
\newblock Provable anonymity.
\newblock In {\em Proc.~3rd ACM Workshop on Formal Methods in Security
  Engineering (FMSE 2005)}, pages 63--72. ACM Press, 2005.

\bibitem{r:gerbrandy99}
J.~Gerbrandy.
\newblock Dynamic epistemic logic.
\newblock In {\em Logic, Language, and Information}, volume~2. CSLI
  Publication, 1999.

\bibitem{r:goguen82}
J.~A. Goguen and J.~Meseguer.
\newblock Security policies and security models.
\newblock In {\em Proc.~1982 IEEE Symposium on Security and Privacy}, pages
  11--20. IEEE Computer Society Press, 1982.

\bibitem{r:goguen84}
J.~A. Goguen and J.~Meseguer.
\newblock Unwinding and inference control.
\newblock In {\em Proc.~1984 IEEE Symposium on Security and Privacy}, pages
  75--86. IEEE Computer Society Press, 1984.

\bibitem{r:goldreich01}
O.~Goldreich.
\newblock {\em Foundations of Cryptography: Volume 1, Basic Tools}.
\newblock Cambridge University Press, 2001.

\bibitem{r:goldreich04}
O.~Goldreich.
\newblock {\em Foundations of Cryptography: Volume 2, Basic Applications}.
\newblock Cambridge University Press, 2004.

\bibitem{r:goldwasser89}
S.~Goldwasser, S.~Micali, and C.~Rackoff.
\newblock The knowledge complexity of interactive proof systems.
\newblock {\em SIAM Journal on Computing}, 18(1):186--208, 1989.

\bibitem{r:gong90}
L.~Gong, R.~Needham, and R.~Yahalom.
\newblock Reasoning about belief in cryptographic protocols.
\newblock In {\em Proc.~1990 IEEE Symposium on Security and Privacy}, pages
  234--248. IEEE Computer Society Press, 1990.

\bibitem{r:gordon03}
A.~D. Gordon and A.~Jeffrey.
\newblock Authenticity by typing for security protocols.
\newblock {\em Journal of Computer Security}, 11(4):451--520, 2003.

\bibitem{r:gray98}
J.~W. Gray, III and P.~F. Syverson.
\newblock A logical approach to multilevel security of probabilistic systems.
\newblock {\em Distributed Computing}, 11(2):73--90, 1998.

\bibitem{r:halpern99}
J.~Y. Halpern.
\newblock Set-theoretic completeness for epistemic and conditional logic.
\newblock {\em Annals of Mathematics and Artificial Intelligence}, 26:1--27,
  1999.

\bibitem{r:halpern88}
J.~Y. Halpern, Y.~Moses, and M.~R. Tuttle.
\newblock A knowledge-based analysis of zero knowledge.
\newblock In {\em Proc.~20th Annual ACM Symposium on the Theory of Computing
  (STOC'88)}, pages 132--147, 1988.

\bibitem{r:halpern94}
J.~Y. Halpern, Y.~Moses, and M.~Y. Vardi.
\newblock Algorithmic knowledge.
\newblock In {\em Proc.~5th Conference on Theoretical Aspects of Reasoning
  about Knowledge (TARK'94)}, pages 255--266. Morgan Kaufmann, 1994.

\bibitem{r:halpern05d}
J.~Y. Halpern and K.~O'Neill.
\newblock Anonymity and information hiding in multiagent systems.
\newblock {\em Journal of Computer Security}, 13(3):483--514, 2005.

\bibitem{r:halpern08}
J.~Y. Halpern and K.~O'Neill.
\newblock Secrecy in multiagent systems.
\newblock {\em ACM Transactions on Information and System Security}, 12(1),
  2008.

\bibitem{r:halpern03d}
J.~Y. Halpern and R.~Pucella.
\newblock On the relationship between strand spaces and multi-agent systems.
\newblock {\em ACM Transactions on Information and System Security},
  6(1):43--70, 2003.

\bibitem{r:halpern11}
J.~Y. Halpern and R.~Pucella.
\newblock Dealing with logical omniscience: Expressiveness and pragmatics.
\newblock {\em Artificial Intelligence}, 175(1):220--235, 2011.

\bibitem{r:halpern12}
J.~Y. Halpern and R.~Pucella.
\newblock Modeling adversaries in a logic for security protocol analysis.
\newblock {\em Logical Methods in Computer Science}, 8(1:21), 2012.

\bibitem{r:halpern93}
J.~Y. Halpern and M.~R. Tuttle.
\newblock Knowledge, probability, and adversaries.
\newblock {\em Journal of the ACM}, 40(4):917--962, 1993.

\bibitem{r:heintze98}
N.~Heintze and J.~G. Riecke.
\newblock The {SLam} calculus: Programming with secrecy and integrity.
\newblock In {\em Proc.~25th Annual ACM Symposium on Principles of Programming
  Languages (POPL'98)}, pages 365--377. ACM Press, 1998.

\bibitem{r:hintikka62}
J.~Hintikka.
\newblock {\em Knowledge and Belief}.
\newblock Cornell University Press, 1962.

\bibitem{r:hintikka75}
J.~Hintikka.
\newblock Impossible possible worlds vindicated.
\newblock {\em Journal of Philosophical Logic}, 4:475--484, 1975.

\bibitem{r:hommersom04}
A.~Hommersom, J.-J. Meyer, and E.~{de Vink}.
\newblock Update semantics of security protocols.
\newblock {\em Synthese}, 142(2):229--267, 2004.

\bibitem{r:honda00}
K.~Honda, V.~Vasconcelos, and N.~Yoshida.
\newblock Secure information flow as typed process behaviour.
\newblock In {\em European Symposium on Programming}, volume 1782 of {\em
  Lecture Notes in Computer Science}, pages 180--199. Springer-Verlag, 2000.

\bibitem{r:meyden10}
X.~Huang and R.~van~der Meyden.
\newblock The complexity of epistemic model checking: Clock semantics and
  branching time.
\newblock In {\em Proc.~19th European Conference on Artificial Intelligence
  (ECAI'10)}, pages 549--554. IOS Press, 2010.

\bibitem{r:hughes04}
D.~Hughes and V.~Shmatikov.
\newblock Information hiding, anonymity and privacy: A modular approach.
\newblock {\em Journal of Computer Security}, 12(1):3--36, 2004.

\bibitem{r:hutter04}
D.~Hutter and A.~Schairer.
\newblock Possibilistic information flow control in the presence of encrypted
  communication.
\newblock In {\em Proc.~9th European Symposium on Research in Computer Security
  (ESORICS'04)}, volume 3193 of {\em Lecture Notes in Computer Science}, pages
  209--224. Springer-Verlag, 2004.

\bibitem{r:konolige86}
K.~Konolige.
\newblock {\em A Deduction Model of Belief}.
\newblock Morgan Kaufmann, 1986.

\bibitem{r:kusters09}
R.~K{\"u}sters and T.~Truderung.
\newblock An epistemic approach to coercion-resistance for electronic voting
  protocols.
\newblock In {\em Proc.~2009 IEEE Symposium on Security and Privacy}, pages
  251--266. IEEE Computer Society Press, 2009.

\bibitem{r:bell73b}
L.~J. LaPadula and D.~E. Bell.
\newblock Secure computer systems: A mathematical model.
\newblock Technical Report MTR-2547, Volume 2, MITRE Corporation, 1973.

\bibitem{r:laud03}
P.~Laud.
\newblock Handling encryption in an analysis for secure information flow.
\newblock In {\em Proc.~12th European Symposium on Programming}, volume 2618 of
  {\em Lecture Notes in Computer Science}. Springer-Verlag, 2003.

\bibitem{r:lincoln98}
P.~Lincoln, J.~C. Mitchell, M.~Mitchell, and A.~Scedrov.
\newblock A probabilistic poly-time framework for protocol analysis.
\newblock In {\em Proc.~5th ACM Conference on Computer and Communications
  Security (CCS'98)}, pages 112--121, 1998.

\bibitem{r:lomuscio06}
A.~Lomuscio and B.~Wo{\'z}na.
\newblock A complete and decidable security-specialised logic and its
  application to the tesla protocol.
\newblock In {\em Proc.~5th International Joint Conference on Autonomous Agents
  and Multiagent Systems (AAMAS'06)}, pages 145--152, 2006.

\bibitem{r:lowe95}
G.~Lowe.
\newblock An attack on the {Needham-Schroeder} public-key authentication
  protocol.
\newblock {\em Information Processing Letters}, 56:131--133, 1995.

\bibitem{r:lowe96}
G.~Lowe.
\newblock Breaking and fixing the {N}eedham-{S}chroeder public-key protocol
  using {FDR}.
\newblock In {\em Proc.~2nd International Workshop on Tools and Algorithms for
  the Construction and Analysis of Systems (TACAS'96)}, volume 1055, pages
  147--166. Springer-Verlag, 1996.

\bibitem{r:lowe98}
G.~Lowe.
\newblock Casper: A compiler for the analysis of security protocols.
\newblock {\em Journal of Computer Security}, 6:53--84, 1998.

\bibitem{r:lowe02}
G.~Lowe.
\newblock Analysing protocols subject to guessing attacks.
\newblock In {\em Proc.~Workshop on Issues in the Theory of Security
  (WITS'02)}, 2002.

\bibitem{r:mantel00}
H.~Mantel.
\newblock Possibilistic definitions of security --- an assembly kit.
\newblock In {\em Proc.~13th IEEE Computer Security Foundations Workshop
  (CSFW'00)}, pages 185--199. IEEE Computer Society Press, 2000.

\bibitem{r:mccullough87}
D.~McCullough.
\newblock Specifications for multi-level security and a hook-up property.
\newblock In {\em Proc.~1987 IEEE Symposium on Security and Privacy}, pages
  161--166. IEEE Computer Society Press, 1987.

\bibitem{r:mccullough88}
D.~McCullough.
\newblock Noninterference and the composability of security properties.
\newblock In {\em Proc.~1988 IEEE Symposium on Security and Privacy}, pages
  177--186. IEEE Computer Society Press, 1988.

\bibitem{r:mclean94}
J.~McLean.
\newblock A general theory of composition for trace sets closed under selective
  interleaving functions.
\newblock In {\em Proc.~1994 IEEE Symposium on Security and Privacy}, pages
  79--93, 1994.

\bibitem{r:meadows95}
C.~Meadows.
\newblock Formal verification of cryptographic protocols: A survey.
\newblock In {\em Advances in Cryptology (ASIACRYPT'94)}, volume 917 of {\em
  Lecture Notes in Computer Science}, pages 133--150. Springer-Verlag, 1995.

\bibitem{r:meadows96}
C.~Meadows.
\newblock The {NRL} protocol analyzer: An overview.
\newblock {\em Journal of Logic Programming}, 26(2):113--131, 1996.

\bibitem{r:merritt83}
M.~J. Merritt.
\newblock {\em Cryptographic Protocols}.
\newblock PhD thesis, Georgia Institute of Technology, 1983.

\bibitem{r:meyer95}
J.-J.~C. Meyer and W.~van~der Hoek.
\newblock {\em Epistemic Logic for {AI} and Computer Science}, volume~41 of
  {\em Cambridge Tracts in Theoretical Computer Science}.
\newblock Cambridge University Press, 1995.

\bibitem{r:micciancio04}
D.~Micciancio and B.~Warinschi.
\newblock Soundness of formal encryption in the presence of active adversaries.
\newblock In {\em Proc.~Theory of Cryptography Conference (TCC'04)}, volume
  2951 of {\em Lecture Notes in Computer Science}, pages 133--151.
  Springer-Verlag, 2004.

\bibitem{r:millen03}
J.~Millen and V.~Shmatikov.
\newblock Symbolic protocol analysis with products and {Diffie-Hellman}
  exponentiation.
\newblock In {\em Proc.~16th IEEE Computer Security Foundations Workshop
  (CSFW'03)}, pages 47--61. IEEE Computer Society Press, 2003.

\bibitem{r:mitchell01}
J.~Mitchell, A.~Ramanathan, A.~Scedrov, and V.~Teague.
\newblock A probabilistic polynomial-time calculus for analysis of
  cryptographic protocols.
\newblock In {\em Proc.~17th Annual Conference on the Mathematical Foundations
  of Programming Semantics}, volume~45 of {\em Electronic Notes in Theoretical
  Computer Science}. Elsevier Science Publishers, 2001.

\bibitem{r:moser90}
L.~E. Moser.
\newblock A logic of knowledge and belief for reasoning about computer
  security.
\newblock In {\em Proc.~3rd IEEE Computer Security Foundations Workshop
  (CSFW'90)}, pages 57--63. IEEE Computer Society Press, 1990.

\bibitem{r:moses88}
Y.~Moses.
\newblock Resource-bounded knowledge.
\newblock In {\em Proc.~2nd Conference on Theoretical Aspects of Reasoning
  about Knowledge (TARK'88)}, pages 261--276. Morgan Kaufmann, 1988.

\bibitem{r:needham78}
R.~M. Needham and M.~D. Schroeder.
\newblock Using encryption for authentication in large networks of computers.
\newblock {\em Communications of the ACM}, 21(12):993--999, 1978.

\bibitem{r:nessett90}
D.~M. Nessett.
\newblock A critique of the {Burrows}, {Abadi} and {Needham} logic.
\newblock {\em ACM Operating Systems Review}, 24(2):35--38, 1990.

\bibitem{r:oHalloran90}
C.~O'Halloran.
\newblock A calculus of information flow.
\newblock In {\em Proc.~European Symposium on Research in Computer Security},
  1990.

\bibitem{r:paulson94}
L.~C. Paulson.
\newblock {\em Isabelle, A Generic Theorem Prover}, volume 828 of {\em Lecture
  Notes in Computer Science}.
\newblock Springer-Verlag, 1994.

\bibitem{r:paulson98}
L.~C. Paulson.
\newblock The inductive approach to verifying cryptographic protocols.
\newblock {\em Journal of Computer Security}, 6(1/2):85--128, 1998.

\bibitem{r:pucella06c}
R.~Pucella.
\newblock Deductive algorithmic knowledge.
\newblock {\em Journal of Logic and Computation}, 16(2):287--309, 2006.

\bibitem{r:raimondi04}
F.~Raimondi and A.~Lomuscio.
\newblock Verification of multiagent systems via ordered binary decision
  diagrams.
\newblock In {\em Proc.~AAMAS'04}, pages 630--637. IEEE Computer Society Press,
  2004.

\bibitem{r:reiter98}
M.~K. Reiter and A.~D. Rubin.
\newblock Crowds: Anonymity for web transactions.
\newblock {\em ACM Transactions on Information and System Security},
  1(1):66--92, 1998.

\bibitem{r:ryan00}
P.~Ryan and S.~Schneider.
\newblock {\em Modelling and Analysis of Security Protocols}.
\newblock Addison Wesley, 2000.

\bibitem{r:ryan98}
P.~Y.~A. Ryan and S.~A. Schneider.
\newblock An attack on a recursive authentication protocol: A cautionary tale.
\newblock {\em Information Processing Letters}, 65(1):7--10, 1998.

\bibitem{r:sabelfeld03}
A.~Sabelfeld and A.~C. Myers.
\newblock Language-based information-flow security.
\newblock {\em IEEE Journal on Selected Areas in Communications}, 21(1), 2003.

\bibitem{r:sabelfeld09}
A.~Sabelfeld and D.~Sands.
\newblock Declassification: Dimensions and principles.
\newblock {\em Journal of Computer Security}, 17(5):517--548, 2009.

\bibitem{r:schneider11}
F.~B. Schneider, K.~Walsh, and E.~G. Sirer.
\newblock Nexus {A}uthorization {L}ogic ({NAL}): Design rationale and
  applications.
\newblock {\em ACM Transactions on Information and System Security},
  14(1):1--28, 2011.

\bibitem{r:schneier96}
B.~Schneier.
\newblock {\em Applied Cryptography}.
\newblock John Wiley \& Sons, second edition, 1996.

\bibitem{r:smith09}
G.~Smith.
\newblock On the foundations of quantitative information flow.
\newblock In {\em Proc.~12th International Conference on Foundations of
  Software Science and Computation Structures (FOSSACS'09)}, volume 5504 of
  {\em Lecture Notes in Computer Science}, pages 288--302. Springer-Verlag,
  2009.

\bibitem{r:smith98}
G.~Smith and D.~Volpano.
\newblock Secure information flow in a multi-threaded imperative language.
\newblock In {\em Proc.~25th Annual ACM Symposium on Principles of Programming
  Languages (POPL'98)}, pages 355--364. ACM Press, 1998.

\bibitem{r:stinson95}
D.~R. Stinson.
\newblock {\em Cryptography: Theory and Practice}.
\newblock CRC Press, 1995.

\bibitem{r:stubblebine96}
S.~Stubblebine and R.~Wright.
\newblock An authentication logic supporting synchronization, revocation, and
  recency.
\newblock In {\em Proc.~3rd ACM Conference on Computer and Communications
  Security (CCS'96)}. ACM Press, 1996.

\bibitem{r:sutherland86}
D.~Sutherland.
\newblock A model of information.
\newblock In {\em Proc.~9th National Computer Security Conference}, pages
  175--183, 1986.

\bibitem{r:syverson90}
P.~Syverson.
\newblock A logic for the analysis of cryptographic protocols.
\newblock NRL Report 9305, Naval Research Laboratory, 1990.

\bibitem{r:syverson99}
P.~Syverson.
\newblock Towards a strand semantics for authentication logic.
\newblock In {\em Proc.~15th Annual Conference on the Mathematical Foundations
  of Programming Semantics}, volume~20 of {\em Electronic Notes in Theoretical
  Computer Science}. Elsevier Science Publishers, 1999.

\bibitem{r:syverson99a}
P.~F. Syverson and S.~G. Stubblebine.
\newblock Group principals and the formalization of anonymity.
\newblock In {\em Proc.~World Congress on Formal Methods in the Development of
  Computing Systems}, volume 1708 of {\em Lecture Notes in Computer Science},
  pages 814--833, 1999.

\bibitem{r:syverson94}
P.~F. Syverson and P.~C. van Oorschot.
\newblock On unifying some cryptographic protocol logics.
\newblock In {\em Proc.~1994 IEEE Symposium on Security and Privacy}, pages
  14--28. IEEE Computer Society Press, 1994.

\bibitem{r:thayer99}
F.~J. Thayer, J.~C. Herzog, and J.~D. Guttman.
\newblock Strand spaces: Proving security protocols correct.
\newblock {\em Journal of Computer Security}, 7(2/3):191--230, 1999.

\bibitem{r:toninho09}
B.~Toninho and L.~Caires.
\newblock A spatial-epistemic logic and tool for reasoning about security
  protocols.
\newblock Technical report, Departamento de Inform{\'a}tica, FCT/UNL, 2009.

\bibitem{r:toussaint89}
M.-J. Toussaint and P.~Wolper.
\newblock Reasoning about cryptographic protocols.
\newblock In J.~Feigenbaum and M.~Merritt, editors, {\em Distributed Computing
  and Cryptography}, volume~2 of {\em DIMACS Series in Discrete Mathematics and
  Theoretical Computer Science}, pages 245--262. American Mathematical Society,
  1989.

\bibitem{r:meyden99}
R.~van~der Meyden and N.~V. Shilov.
\newblock Model checking knowledge and time in systems with perfect recall
  (extended abstract).
\newblock In {\em Proc.~Conference on Foundations of Software Technology and
  Theoretical Computer Science}, volume 1738 of {\em Lecture Notes in Computer
  Science}, pages 432--445. Springer-Verlag, 1999.

\bibitem{r:meyden04}
R.~van~der Meyden and K.~Su.
\newblock Symbolic model checking the knowledge of the dining cryptographers.
\newblock In {\em Proc.~17th IEEE Computer Security Foundations Workshop
  (CSFW'04)}, pages 280--291. IEEE Computer Society Press, 2004.

\bibitem{r:ditmarsch03}
H.~P. van Ditmarsch.
\newblock The {Russian} {Cards} problem.
\newblock {\em Studia Logica}, 75:31--62, 2003.

\bibitem{r:ditmarsch06}
H.~P. van Ditmarsch, W.~van~der Hoek, R.~van~der Meyden, and J.~Ruan.
\newblock Model checking russian cards.
\newblock {\em Electronic Notes in Theoretical Computer Science},
  149(2):105--123, 2006.

\bibitem{r:vaneijck07}
J.~van Eijck and S.~Orzan.
\newblock Epistemic verification of anonymity.
\newblock {\em Electronic Notes in Theoretical Computer Science}, 168:159--174,
  2007.

\bibitem{r:oorschot93a}
P.~C. van Oorschot.
\newblock Extending cryptographic logics of belief to key agreement protocols.
\newblock In {\em Proc.~1st ACM Conference on Computer and Communications
  Security (CCS'93)}, pages 232--243. ACM Press, 1993.

\bibitem{r:wedel96}
G.~Wedel and V.~Kessler.
\newblock Formal semantics for authentication logics.
\newblock In {\em Proc.~4th European Symposium on Research in Computer Security
  (ESORICS'96)}, volume 1146 of {\em Lecture Notes in Computer Science}, pages
  219--241. Springer-Verlag, 1996.

\bibitem{r:wittbold90}
J.~T. Wittbold and D.~Johnson.
\newblock Information flow in nondeterministic systems.
\newblock In {\em Proc.~1990 IEEE Symposium on Security and Privacy}. IEEE
  Computer Society Press, 1990.

\end{thebibliography}

\end{document}